\documentclass[5p,sort&compress]{elsarticle}

\usepackage{mathtools}

\usepackage{siunitx}

\usepackage[dvipsnames]{xcolor}
\usepackage{graphicx}

\usepackage{soul} 
\usepackage[colorlinks=true]{hyperref}

\usepackage{subcaption}

\usepackage[markup=underlined]{changes}

\newcommand{\iu}{\mathrm{i}}
\newcommand{\vect}[1]{\boldsymbol{#1}}

\journal{Acta Acustica}

\begin{document}

\begin{frontmatter}

\title{\textbf{Effects of the sheared flow velocity profile\\on impedance eduction in a 2D duct}}

\def\elspublication{Published in Acta Acoustica (2026), vol.~10, 2026005, \href{https://doi.org/10.1051/aacus/2026005}{doi:10.1051/aacus/2026005}}

\author[label1,label2]{Lucas~Araujo~Bonomo}\ead{lucas.bonomo@lva.ufsc.br}
\author[label2,label3]{Edward~James~Brambley}\ead{E.J.Brambley@warwick.ac.uk}
\author[label1]{Julio~Apolinário~Cordioli}\ead{julio.cordioli@ufsc.br}

\affiliation[label1]{organization={Department of Mechanical Engineering},
            addressline={Federal University of Santa Catarina}, 
            city={Florianópolis},
            postcode={88040-900}, 
            state={SC},
            country={Brazil}}

\affiliation[label2]{organization={Mathematics Institute},
            addressline={University of Warwick}, 
            city={Coventry},
            postcode={CV4 7AL}, 
            country={United Kingdom}}
            
\affiliation[label3]{organization={WMG},
            addressline={University of Warwick}, 
            city={Coventry},
            postcode={CV4 7AL}, 
            country={United Kingdom}}

\begin{abstract}
    Impedance eduction methods are the current standard approach to measure the impedance of acoustic liner under sheared grazing flow. The dedicated facilities for these methods consists on a waveguide with rectangular cross-section, which implies a sheared grazing flow. A current debate in the literature is the effect of this sheared flow in the impedance eduction methods. We assess the impact of the flow profile shape on acoustic propagation in a two-dimensional duct within the typical operating range of impedance eduction facilities. Firstly, a numerical experiment is proposed in which the Pridmore--Brown equation is assumed to represent the true physical behaviour, and is used with both simplified flow profiles commonly used in the literature and a realistic representation of a turbulent boundary layer using a van Driest universal law of the wall model. The data from these numerical experiments are then used with a traditional impedance eduction process, and the resulting variation in obtained impedances are investigated.  Secondly, we apply a less-traditional impedance eduction method that incorporates the sheared velocity profile to data obtained from real-world experiments. The results suggest that the Ingard--Myers boundary condition remains a good approximation  to a realistic boundary layer profile, such as the universal law of the wall, at least in the two-dimensional case. However, it is also shown that the simplified flow profiles often used in the literature can lead to significant deviations from the results obtained using a realistic velocity distribution.
\end{abstract}


\end{frontmatter}

\section{Introduction}

While the ERI method allows to localize defects that
are not in direct view from the transducers, it is useful
to also examine its performance for a defect in direct
view. Numerical experiments are performed in the same
reference medium, with a local density anomaly inserted
this time in front of the transducer array. Corresponding 
images are presented in Figure 6 for different window
durations.

    Acoustic liners are acoustic treatments applied to the walls of aircraft turbofan engine nacelles to mitigate fan noise. The simplest and most typical liner construction consists of a honeycomb structure with a hard backplate and a perforated facesheet~\citep{smith1989aircraft}. An acoustic liner is typically characterised by its locally-reacting acoustic impedance, $\Tilde{Z}(\omega) = \theta + \iu\chi$, where $\theta$ is the resistance and $\chi$ is the reactance. This frequency-dependent parameter can be used as a boundary condition in simulations of aircraft engine noise, avoiding the still prohibitive computational cost of explicitly modelling an acoustic liner. 

    The impedance of an acoustic liner is known to depend on its geometry~\citep{guess1975Calculation}, as well as on operational conditions such as the grazing flow velocity and profile~\citep{kooi1981Experimental}, and the incident Sound Pressure Level (SPL)~\citep{murray2012development}. Therefore, for proper liner characterisation, experiments must replicate the conditions inside a turbofan engine, and this is traditionally carried out using the so-called in-situ technique (or Dean's method) \citep{dean1974insitu} or impedance eduction methods \citep{watson1999validation, jing2008straightforward, elnady2009validation}. While the in-situ technique provides a local value of the liner impedance, impedance eduction techniques give an averaged impedance as seen by the acoustic field and are experimentally simpler to implement than the in-situ technique. As a result, impedance eduction has been preferred by the academic community \citep[e.g.][]{ferrante2016back, bonomoComparisonSituImpedance2024}. Impedance eduction methods rely on a model for acoustic propagation in ducts, inferring the impedance from the model by best fitting the model results to the experimental data.  Uniform flow and the Ingard--Myers boundary condition are the most common modelling assumptions. However, recent findings have sparked academic debate regarding the application of such assumptions.

    One of the fundamental issues recently observed is that educed impedance depends on the direction of wave propagation relative to the mean flow, which constitutes a violation of the locally reacting hypothesis. This behaviour has been systematically captured by different laboratory facilities using both inverse and direct methods, as summarised by \citet{boden2016Effect}. \citet{renou2011Failure} were the first to demonstrate such discrepancies and attributed them to a failure of the Ingard--Myers boundary condition. Since then, other studies have suggested that an additional parameter, other than the impedance, is necessary to fully characterise acoustic liners in the presence of flow, such as a shear stress at the wall~\citep{auregan2018stress, weng2018Flow}, or the effect of viscosity~\citep{khamis2017acoustics}.
    
    Another line of research involves substituting the traditional uniform flow hypothesis with a shear flow profile. Despite the three-dimensional nature of internal flows, most authors start by making a simplification and consider a two-dimensional representation of the computational domain, which implies a one-dimensional flow profile \mbox{\citep{primus2013ONERANASA, jingInvestigationStraightforwardImpedance2015, weng2018Flow, yangShearFlowEffects2024}}.
    The classical analysis by \mbox{\citet{pridmorebrown1958sound}} established the formulation of sound propagation in ducts with sheared flow, providing the theoretical foundation for subsequent developments. Later, \mbox{\citet{brooks2007Sound}} extended this framework to examine the transmission of sound in ducts with realistic turbulent shear layers, focusing on the dispersion and attenuation characteristics of the acoustic modes.
    While \citet{weng2018Flow} and \citet{yangShearFlowEffects2024} showed that shear effects become increasingly relevant at higher frequencies and larger Helmholtz numbers, many studies still rely on simplified flow profiles for computational convenience. \citet{roncen2020WavenumberBased} revealed significant differences in wavenumbers between uniform flow solutions and those using two-dimensional sheared profiles, indicating that simplifications can introduce bias errors related to the upstream-downstream impedance discrepancy. However, despite highlighting these discrepancies, \citet{roncen2020WavenumberBased}’s subsequent impedance eduction still used a simplified one-dimensional profile, potentially undermining the accuracy of their conclusions. Since the early work of \citet{nayfeh1974Effect}, it has been generally assumed that matching the boundary layer displacement thickness is sufficient for predicting acoustic behaviour in downstream propagation; however, \citet{nayfeh1974Effect} also showed that for upstream propagation, the flow shape factor can significantly influence sound attenuation and wave propagation.
    \mbox{\citet{gabard2013Comparison}} also investigated the effect of sheared velocity profiles, but in contrast, concluded that while ``the boundary layer thickness can have a significant impact on sound absorption, \ldots\ the boundary layer profile is found to have little influence on sound absorption.''
    All of these studies primarily addressed propagation and absorption in large-scale ducts or benchmark configurations, rather than the inverse problem of impedance eduction in small, plane-wave ducts typical of laboratory experiments.
    
    More recently, \mbox{\citet{weng2018Comparison}} compared impedance eduction techniques under uniform and sheared flow conditions, using a measured turbulent profile representative of their experimental setup; they focused on a single measured profile, and did not investigate how different boundary-layer shapes might affect impedance eduction outcomes.

     The present work builds upon these foundations by explicitly quantifying how the shape of the velocity profile—modelled through sinusoidal, hyperbolic tangent, and van Driest formulations—affects impedance eduction results obtained from both direct and inverse methods. By systematically varying the flow-profile shape under controlled numerical and experimental conditions, the study isolates the role of the boundary-layer distribution in the impedance eduction process, thereby extending the understanding of shear-flow effects to the low-Mach-number, small-duct regime relevant to experimental liner characterisation.
    This effect may help explain if the assumption of a uniform flow has a role in the current debate regarding the upstream and downstream discrepancy. 
    In order to achieve this goal, a numerical experiment approach is proposed. The solution of the Pridmore--Brown equation~\citep{pridmorebrown1958sound} is assumed as the exact solution for the acoustic field propagating over a sheared mean flow, from which the axial wavenumbers in an infinite two-dimensional duct are obtained. These wavenumbers are then used in the traditional straightforward impedance eduction routine~\citep{jing2008straightforward}, assuming uniform flow and the Ingard--Myers boundary condition. In addition, a parametric study is conducted within the scope of this numerical experiment to evaluate the impact of the test rig duct width and the average Mach number on the accuracy of the Ingard--Myers boundary condition. Finally, we employ an iterative impedance eduction method on experimental data, similar to the one used by \citet{roncen2020WavenumberBased}, to analyse the impact of different flow velocity profile distributions on impedance eduction, comparing these results with the estimates provided by the Ingard--Myers boundary condition.

    This document is organised as follows. Section~\ref{sec:equations} presents the governing equations for duct acoustics with grazing flow. Section~\ref{sec:flow} describes the flow velocity distributions considered. The setup for the numerical experiments is detailed in Section~\ref{sec:materials}. The main theoretical results are discussed in Section~\ref{sec:results}, while application to experimental data is presented in Section~\ref{sec:experiments}. Finally, the primary conclusions are summarised in Section~\ref{sec:conclu}.

\section{Governing equations} \label{sec:equations}

\begin{figure}[htb!]
        \centering%
        \includegraphics[width=\linewidth]{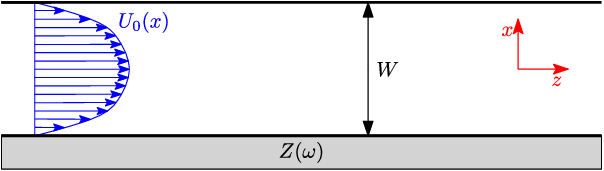}%
        \caption{Schematic duct and coordinates system assumed in this work.}%
        \label{fig:schematicDuct}%
    \end{figure}

    For the purpose of this study, we consider the infinite 2D duct depicted in Fig.~\ref{fig:schematicDuct}.
    The duct cross-section has width $W$. An axial flow with velocities profile $\vect{u}_0 = U_0(x)\vect{\hat{\mathrm{k}}}$ is assumed, where $\vect{\hat{\mathrm{k}}}$ is the unitary vector in $z$ axis, which implies that the flow profile has no dependence on the axial direction. The wall located at $x = -W/2$ has a locally-reacting frequency-dependent impedance $Z(\omega)$, while the other wall at $x=W/2$ is acoustically rigid.    

     By considering linear perturbations and constant mean density and sound speed, while neglecting visco-thermal effects, the in-duct acoustic propagation can be described by the Pridmore–Brown equation (PBE)~\citep{pridmorebrown1958sound}, such that
    \begin{multline}
        (\iu \omega + \vect{u_0\cdot\nabla}) \left( \dfrac{1}{c_0^2} (\iu \omega + \vect{u_0 \cdot \nabla})^2 \tilde{p}' - \nabla^2 \tilde{p}' \right)
        \\
        + 2 \dfrac{\partial}{\partial z} \big(\vect{\nabla} \tilde{p}' \vect{\cdot \nabla} U_0\big) = 0, \label{eq:PB_generalForm}
    \end{multline}
    where $\tilde{p}'$ is the acoustic pressure, $\omega$ is the frequency ($\exp(\iu\omega t)$ dependence assumed), $c_0$ the speed of sound, $\iu = \sqrt{-1}$ the complex imaginary unity and $\nabla = (\partial/\partial x, \partial/\partial z)$. Given the axial invariance of the problem, we can assume an axial modal solutions on the form $\tilde{p}'(x,z) = \tilde{p}'(x) \exp(-\iu k_z z)$, where $k_z$ is the axial wavenumber, so that Eq.~\eqref{eq:PB_generalForm} can be written as
    \begin{multline}
        \!\left(\!\nabla_\perp^2 + \dfrac{\omega^2}{c_0^2}\right)\!\tilde{p}' - k_z \left( \dfrac{U_0}{\omega}  \nabla_\perp^2 - \dfrac{2}{\omega} \nabla_\perp U_0 \cdot \nabla_\perp + \dfrac{3 \omega U_0}{c_0^2} \right)\! \tilde{p}' \\ - k_z^2 \!\left(\! 1 - \dfrac{3U_0^2}{c_0^2} \right)\! \tilde{p}' - k_z^3\! \left[ \dfrac{U_0}{\omega} \!\left( \dfrac{U_0^2}{c_0^2} - 1\!\right)\! \right]\! \tilde{p}' = 0, \label{eq:PBE_kz}
    \end{multline}
    where $\nabla_\perp =  (\partial/\partial x, 0)$. As boundary conditions, at rigid walls the normal acoustic velocity $\mathbf{u}'$ vanishes, such that
    \begin{equation}
        \vect{u'}\cdot\vect{\hat{\mathrm{n}}} = 0, 
    \end{equation}
    where $\vect{\hat{\mathrm{n}}}$ is a unitary normal vector pointing into the wall. Since non-slip flows are assumed, the locally reacting impedance boundary condition can be written as
    \begin{equation}
        Z = \dfrac{1}{\rho_0 c_0} \dfrac{\tilde{p}'}{\vect{u'}\cdot\vect{\hat{\mathrm{n}}}}, \label{eq:Z}
    \end{equation}
    where the air characteristic impedance $\rho_0 c_0$ is used as a normalisation factor and $\rho_0$ is the air density. For the two-dimensional duct assumed in this work, $\vect{\hat{\mathrm{n}}} = \vect{\hat{\mathrm{i}}}$ at $x=W/2$, where $\vect{\hat{\mathrm{i}}}$ is the unitary vector in $x$ axis, and $\vect{\hat{\mathrm{n}}} = -\vect{\hat{\mathrm{i}}}$ at $x=-W/2$. For later convenience, we also introduce the distance to the wall $\xi = W/2 - |x|$.

    \subsection{Eigenvalue problem}

        In this section, we seek to describe the governing equations as a generalized eigenvalue problem. One can rewrite the PBE~(Eq.~\eqref{eq:PBE_kz}) in a discrete version as
        \begin{equation}
            (\mathbf{A}_0 + \mathbf{A}_1 k_z + \mathbf{A}_2 k_z^2 + \mathbf{A}_3 k_z^3) \mathbf{\tilde{p}} = \mathbf{0}, \label{eq:genEig}
        \end{equation}
        where the $\mathbf{A}_j$ terms involve differentiation in $x$ and multiplication by the frequency $\omega$ and the mean flow $U_0$ and its $x$-derivatives. In the present work, we follow a strategy similar to \citet{boyer2011Theoretical}, where the problem is discretised by projecting onto a Gauss--Lobatto grid using Chebyshev polynomials as basis, with~\eqref{eq:genEig} required to hold at each grid point (a pseudo-spectral method). Finally, to solve the cubic generalised eigenvalue problem given by Eq.~\eqref{eq:genEig}, auxiliary variables are introduced of the form $\mathbf{\tilde{p}}_p~=~k_z\mathbf{\tilde{p}}_{p-1}$ for $p>0$, and the resulting linear eigenvalue problem is solved using the QZ algorithm \cite[p. 129]{boyd2001Chebyshev}.

        In order to apply a lined wall boundary condition to the generalised eigenvalue problem, we rewrite Eq.~\eqref{eq:Z} as 
        \begin{equation}
            \dfrac{\mathrm{d} \tilde{p}'}{\mathrm{d} x} n_x + \dfrac{\iu \omega}{c_0 Z}\tilde{p}' = 0, \label{eq:rigid}
        \end{equation}        
        where $n_x = -1$ at $x = -W/2$. For the hard wall opposite to the liner, the corresponding boundary condition is
        \begin{equation}
            \dfrac{\mathrm{d} \tilde{p}'}{\mathrm{d} x} = 0.
        \end{equation}

        If a uniform flow is assumed, i.e.\ $U_0 \equiv M c_0 = \text{constant}$, where $M$ is the bulk (average) Mach number, the PBE reduces to the Convected Helmholtz Equation (CHE),
            \begin{equation}
             \nabla^2 \tilde{p}' + \left(k_0 -\iu M \frac{\partial}{\partial z}\right)^2 \!\!\tilde{p}' = 0. \label{eq:CHE}
            \end{equation}
        where $k_0 \equiv \omega /c_0$ is the free-field wavenumber. For lined walls, the slip velocity at the wall is taken into account by means of the Ingard--Myers Boundary Condition (IMBC)~\citep{ingard1959influence,myers1980acoustic}, leading to
            \begin{equation}
        		\frac{\partial \tilde{p}'}{\partial x} = \frac{1}{\iu k_0 Z} \left( \iu k_0 + M_w \dfrac{\partial}{\partial z} \right)^2 \tilde{p}', \label{eq:IMBC}
            \end{equation}
        where $M_w$ is the slipping velocity at the wall, and so $M_w = M$ for a uniform flow.  For a non-slip flow, $M_w=0$, and Eq.~\eqref{eq:IMBC} reduces to Eq.~\eqref{eq:rigid}.

    \subsection{Impedance eduction} \label{subsec:eduction}

        In this work, we consider the traditional straightforward wavenumber based impedance eduction first proposed by \citet{jing2008straightforward}. Applying the IMBC (Eq.~\eqref{eq:IMBC}) on the lined wall and Eq.~\eqref{eq:rigid} on the rigid walls of the CHE solution leads to the eigenvalue problem
        \begin{equation} \label{eq:eigenvalue}
            k_x \tan(k_x W) - \frac{1}{\iu k_0 Z} \left(\iu k_0 - \iu Mk_z \right)^2 = 0, 
        \end{equation}
        where $k_x$ is the transverse wavenumber given by the dispersion relation
        \begin{equation}
            k_x^2 = \left( k_0 - M k_z \right)^2 - k_z^2. \label{eq:dispRel}
        \end{equation}
        Once the axial wavenumber is known, it is straightforward to calculate the liner impedance from Eqs.~\eqref{eq:eigenvalue} and \eqref{eq:dispRel}.

\section{Velocities profile shape functions} \label{sec:flow}

    The simplest formulation considered in this work is the sinusoidal flow profile, as presented by \mbox{\citet{gabard2013Comparison}}. In this case,  
    \begin{equation}
        \dfrac{U_0({\color{blue}\xi})}{c_0} = \begin{cases}
            M_s \sin \dfrac{\pi \xi}{2\delta_s}, & 0\leq \xi \leq \delta_s\\
            M_s, & \xi > \delta_s,
        \end{cases} \label{eq:sine}
    \end{equation}
    where $M_s$ is the free-stream Mach number, and $\delta_s$ is a boundary layer thickness parameter.  
    
    Another commonly employed formulation in the literature is the hyperbolic tangent profile introduced by \citet{rienstra2008spatial}, and used by \citet{roncen2020WavenumberBased} in their work on 2D flow profile effects on impedance eduction. This profile is given by  
    \begin{align}&
        \dfrac{U_0(r)}{c_0} = M_c\! \left[ \tanh{\!\left(\!\dfrac{1-r}{\delta_t}\right)}\!
        \right.\\\notag&\left.
        + (1-\tanh{(1/\delta_t)})\!\left( \dfrac{1+\tanh{(1/\delta_t)}}{\delta_t}r + (1+r)\!\!\right)\!(1-r)\right]\!, \label{eq:tanh}
    \end{align}
    where $M_c$ is the centreline Mach number, $r$ is the radial position, and $\delta_t$ is a boundary layer thickness parameter. In this work, we use the coordinate transformation $r = 2|x|/W$ to obtain the flow profiles in the $x$ coordinate system.

    We also aim to consider a more realistic representation of a turbulent boundary layer velocity profile. One may express the boundary profile over a smooth wall using a universal wall law, which, according to \citet{van1956turbulent}, is given by
    \begin{equation}
        U^+ =  \int_0^{\mathrlap{y^+}} \dfrac{2}{1 + \sqrt{1 + 4 \kappa^2 {y^+}^2(1-\exp(-y^+/A^+))^2}} \mathrm{d}y^+ + \Pi, \label{eq:vanDriest}
    \end{equation}
    where $U^+ \equiv U_0 / u_\tau$ is the flow profile normalised by the friction velocity $u_\tau$, $\kappa \approx 0.42$ is the von Kármán constant, $A^+ \approx 27$ is the van Driest constant, and $y^+ = \xi u_\tau / \nu$ is the distance from the wall, $\xi$, normalised to viscous lengths, with $\nu$ being the air kinematic viscosity. As will be discussed later, for the small ducts considered in this study, the boundary layer can extend the entire half-duct width. To ensure that the derivative of the profile is continuous at the duct centreline, we propose adding a cubic term to Eq.~\eqref{eq:vanDriest}, denoted by $\Pi$, which is given by
    \begin{equation}
        \Pi = \dfrac{2(y^+_{\text{max}} - y^+)}{1 + \sqrt{1 + 4 \kappa^2 {y^+_{\text{max}}}^2(1-\exp(-y^+_{\text{max}}/A^+))^2}} \left(\dfrac{y^+}{y^+_{\text{max}}}\right)^{\!2}\!\!,\label{eq:Pi}
    \end{equation}
    where $y^+_{\text{max}} = W u_\tau / \nu / 2$ is the distance from the wall to the centreline in viscous lengths\footnote{Another option would be the application of blending functions, as has been done by \mbox{\citet{yangShearFlowEffects2024}}}. One of the simplest formulations for the boundary layer shape is the assumption of a linear variation. However, for the purposes of this work, the linear flow profile is not advantageous, as, although it is continuous, its first derivative is not continuous, which compromises the convergence of the pseudospectral method used in this study \cite[p. 32]{boyd2001Chebyshev}. Therefore, we will not use it. Another common simplification for the flow profile shape is the inverse power law, which provides a good approximation of a turbulent boundary layer~\citep{nayfeh1974Effect, yangShearFlowEffects2024}. However, the inverse power law formulation has a problem since the velocity gradient near the wall tends to infinity. One could avoid this problem either by solving for acoustic displacement, as done by \citet{yangShearFlowEffects2024}, which implies that the gradient of the mean flow does not appear explicitly in the equation, or by adding a linear sub-layer near the wall~\citep[e.g.][]{nayfeh1974Effect}, which would result in an additional discontinuity in the derivative. Both solutions add complexity to this problem, and neither is as accurate as using the turbulent wall law formulation from Eq.~\eqref{eq:vanDriest}.

    The boundary layer thickness is often quantified by its displacement thickness or momentum thickness. The boundary layer displacement thickness is the distance by which the external flow must be displaced to account for the reduction in flow caused by the boundary layer, and is defined as
    \begin{equation}
        \delta^* = \int_0 ^\infty \left( 1 - \dfrac{U_0(x)}{U_\infty} \right) \mathrm{d}x,
    \end{equation}
    where $U_\infty$ is the free-stream velocity, which for the purpose of this work is $U_\infty = U_0(0)$. The boundary layer momentum thickness is a measure of the loss of momentum within the boundary layer due to viscous effects and is given by
    \begin{equation}
        \theta = \int_0 ^\infty  \dfrac{U_0(x)}{U_\infty} \left( 1 - \dfrac{U_0(x)}{U_\infty} \right) \mathrm{d}x.
    \end{equation}
    Another important quantity is the boundary layer shape factor, which is the ratio of the displacement thickness to the momentum thickness, defining the shape of the velocity profile within a boundary layer, hence given by
    \begin{equation}\label{equ:shape-factor}
        H = \dfrac{\delta^*}{\theta}.
    \end{equation}
    Note that the shape factor is independent of the boundary layer thickness, and so is a property of the boundary layer flow profile's shape only.  For the flow profiles considered in this work, the shape factors are 1.25, 2.34 and 2.66, for the law of the wall, hyperbolic tangent and sinusoidal profile, respectively.

\section{Numerical setup} \label{sec:materials}

    In this work, we consider a small rectangular duct, representative of traditional liner impedance eduction facilities~\citep{renou2011Failure, ferrante2016back, bonomoComparisonSituImpedance2024, weng2018Flow}. Initially, we consider the dimensions of the Liner Impedance Test Rig from the Federal University of Santa Catarina (LITR/UFSC), which has a rectangular cross-section with a width of $W = \SI{40}{\milli\meter}$.

\begin{figure*}%
        \centering%
        \includegraphics{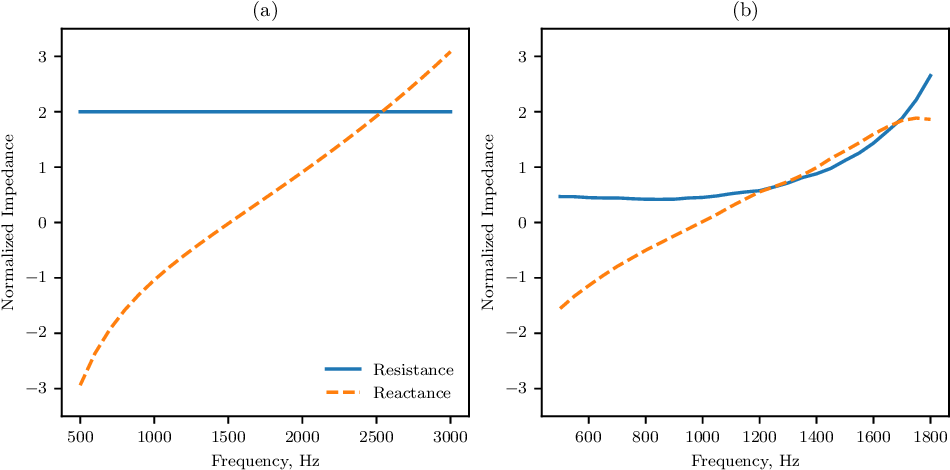}%
        \caption{Reference impedances for numerical experiments. (a) SDOF-like; (b) digitalization of CT57 from \citet{roncen2020WavenumberBased}.}%
        \label{fig:refZ}%
    \end{figure*}

    For the lined wall impedance, two reference impedances are considered. First, we use the impedance given by 
        \begin{equation}
            Z_{\text{SDOF}}(\omega) = 2 - \iu \left( \cot{(k_0 h)} - (0.03k_0)^2 \right),
        \end{equation}
    with $h = \SI{35}{\milli\meter}$, which is representative of a typical Single-Degree-Of-Freedom (SDOF) liner in the considered frequency range, and is shown in Figure~\ref{fig:refZ}a.
    Additionally, one of the reference impedances considered by \citet{roncen2020WavenumberBased} is replicated in this work. The impedance modelling the ceramic liner CT57 was digitized for use in this study and is denoted $Z_{\text{CT57}}$, as presented in Figure~\ref{fig:refZ}b. The typical range for impedance eduction, from 500 to \SI{3000}{\hertz}, with a \SI{50}{\hertz} step, is used for $Z_{\text{SDOF}}$, while a reduced range from 500 to \SI{1800}{\hertz} is used for $Z_{\text{CT57}}$ due to the range of frequencies for which data is available.

    For the velocity profiles, we consider the three formulations presented in Section~\ref{sec:flow}. The turbulent universal wall law, given by Eq.~\eqref{eq:vanDriest}, with $\nu = \SI{1.48e-5}{\meter^2\per\second}$ and $u_\tau = \SI{3.956}{\meter\per\second}$, is selected as the baseline case. This corresponds to the fit of Eq.~\eqref{eq:vanDriest} to experimental data gathered at the LITR/UFSC, allowing for the comparison of different flow profile formulations with a realistic velocity distribution in a typical liner test rig duct. This leads to an average Mach number over the 1D cross-section of $M = \num{0.279}$, a boundary layer thickness of $\delta_{\SI{99}{\percent}} = \SI{15.72}{\milli\meter}$, and a boundary layer displacement thickness of $\delta^* = \SI{1.70}{\milli\meter}$. 
    
\begin{figure*}%
        \centering%
        \includegraphics{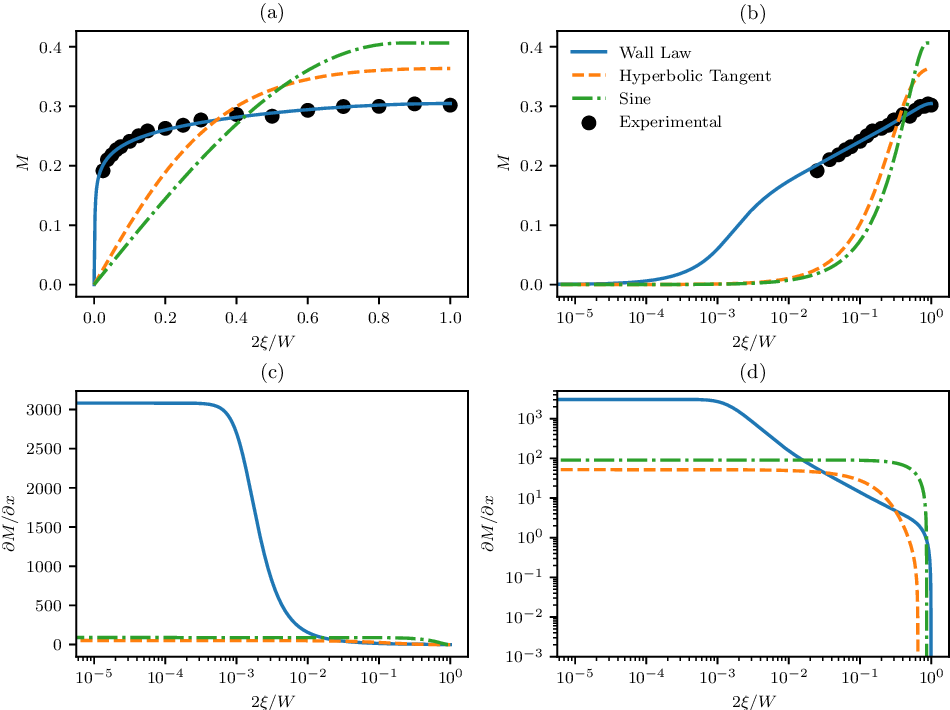}%
        \caption{Flow velocities profiles in linear (a) and logarithmic (b) scales, and velocity gradients in linear (c) and logarithmic (d) scales, considered in the first step of this work. Hyperbolic tangent and sinusoidal flow profiles are best fitted to match the bulk Mach number and boundary layer thickness of the law of the wall.}%
        \label{fig:fitDelta}%
    \end{figure*}

    We aim to reproduce the study of \citet{nayfeh1974Effect} in the context of impedance eduction. To do so, we first need to find the parameters for the hyperbolic tangent and sinusoidal flow profiles that match the same average Mach number $M$ and boundary layer thickness $\delta_{\SI{99.9}{\percent}}$ as the baseline case. This results in different boundary layer displacement thicknesses $\delta^*$ for each flow profile formulation, which is expected to lead to different acoustic fields. The parameters obtained for both the hyperbolic tangent and sinusoidal flow profiles are summarized in Table~\ref{tab:fitDelta},
    \begin{table*}[t]%
        \centering%
        \caption{Resulting parameters for velocities profile formulations fit to baseline case average Mach number and $\delta_{\SI{99}{\percent}}$. Baseline case corresponds to universal wall law with $\nu = \SI{1.48e-5}{\meter^2\per\second}$ and $u_\tau = \SI{3.956}{\meter\per\second}$.}%
        \label{tab:fitDelta}%
        \begin{tabular}{rllcc}
        \hline
        Velocities Profile & \multicolumn{2}{c}{Adjusted Parameters} & Resulting $\delta^*$ & $\delta_{\SI{99}{\percent}}$ \\ \hline
        Hyperbolic Tangent &$M_c = 0.363$ & $\delta_t = 0.3546$& \SI{4.66}{\milli\meter}          & \SI{15.72}{\milli\meter}\\
        Sinusoidal         &$M_s = 0.406 $ & $\delta_s = \SI{17.3}{\milli\meter}$& \SI{6.28}{\milli\meter}     & \SI{15.72}{\milli\meter}     \\ \hline
        \end{tabular}%
    \end{table*}%
    along with the resulting $\delta^*$ for each case. The different flow profiles are compared to the experimental data gathered at the facility in Figure~\ref{fig:fitDelta},
    where the derivative of the velocity distribution is also presented.
    
\begin{figure*}%
        \centering%
        \includegraphics{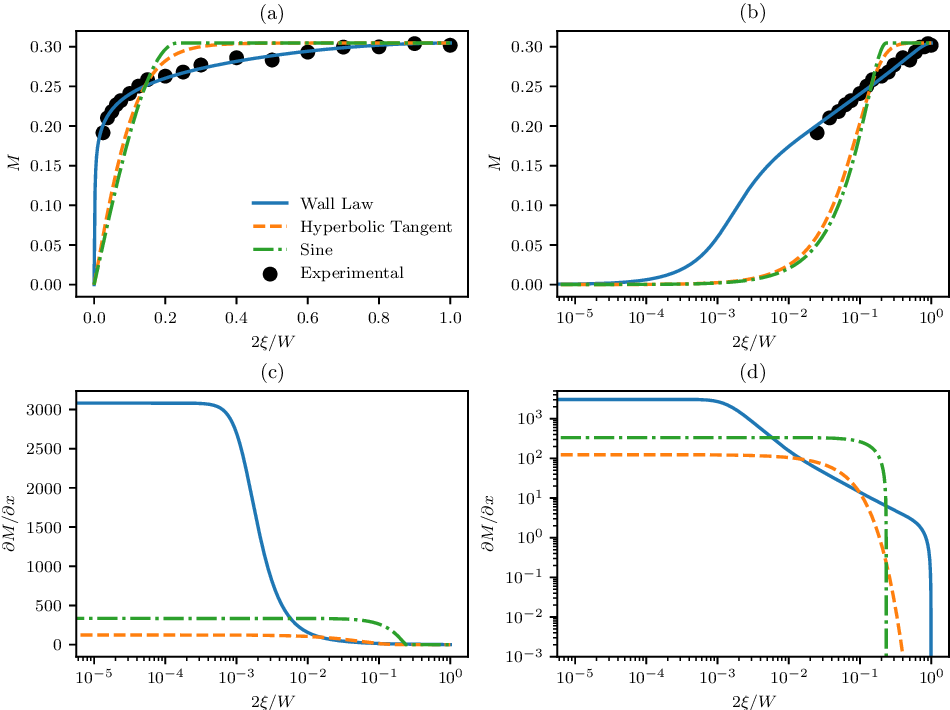}%
        \caption{Flow velocities profiles in linear (a) and logarithmic (b) scales, and velocity gradients in linear (c) and logarithmic (d) scales, considered in the first step of this work. Hyperbolic tangent and sinusoidal flow profiles are best fitted to match the bulk Mach number and boundary layer displacement thickness of the law of the wall.}%
        \label{fig:fitDeltaStar}%
    \end{figure*}

    Next, we consider the case where the different velocity distributions share the same average Mach number and boundary layer displacement thickness $\delta^*$. The parameters obtained for both the hyperbolic tangent and sinusoidal flow profiles under this new condition are summarized in Table~\ref{tab:fitDeltaStar},
    \begin{table*}[t]%
        \centering%
        \caption{Resulting parameters for velocities profile formulations fit to baseline case average Mach number and $\delta^*$. Baseline case corresponds to universal wall law with $\nu = \SI{1.48e-5}{\meter^2\per\second}$ and $u_\tau = \SI{3.75}{\meter\per\second}$.}%
        \label{tab:fitDeltaStar}%
        \begin{tabular}{rllcc}
        \hline
        Velocities Profile & \multicolumn{2}{c}{Adjusted Parameters} & Resulting $\delta_{\SI{99}{\percent}}$ & $\delta^*$ \\ \hline
        Hyperbolic Tangent & $M_c = 0.305$ & $\delta_t = 0.1227$ &  \SI{6.50}{\milli\meter} &     \SI{1.70}{\milli\meter}  \\
        Sinusoidal         & $M_s = 0.305$ & $\delta_s = \SI{4.7}{\milli\meter}$ & \SI{4.26}{\milli\meter}    &    \SI{1.70}{\milli\meter}   \\ \hline
        \end{tabular}%
    \end{table*}%
    along with the resulting $\delta_{\SI{99}{\percent}}$. The resulting flow profiles are once again compared to the experimental data in Figure~\ref{fig:fitDeltaStar}.

    The number of points in the computational domain used for the pseudospectral solver was determined based on the critical case, which, for this study, corresponds to the universal wall law due to its high gradient near the walls. The assessment of the convergence of the numerical grid is detailed in \mbox{\ref{ap:convergence}}.

\section{Theoretical results and discussion} \label{sec:results}

\subsection{Effects of flow profiles on axial wavenumbers}

\begin{figure*}%
        \centering%
        \includegraphics{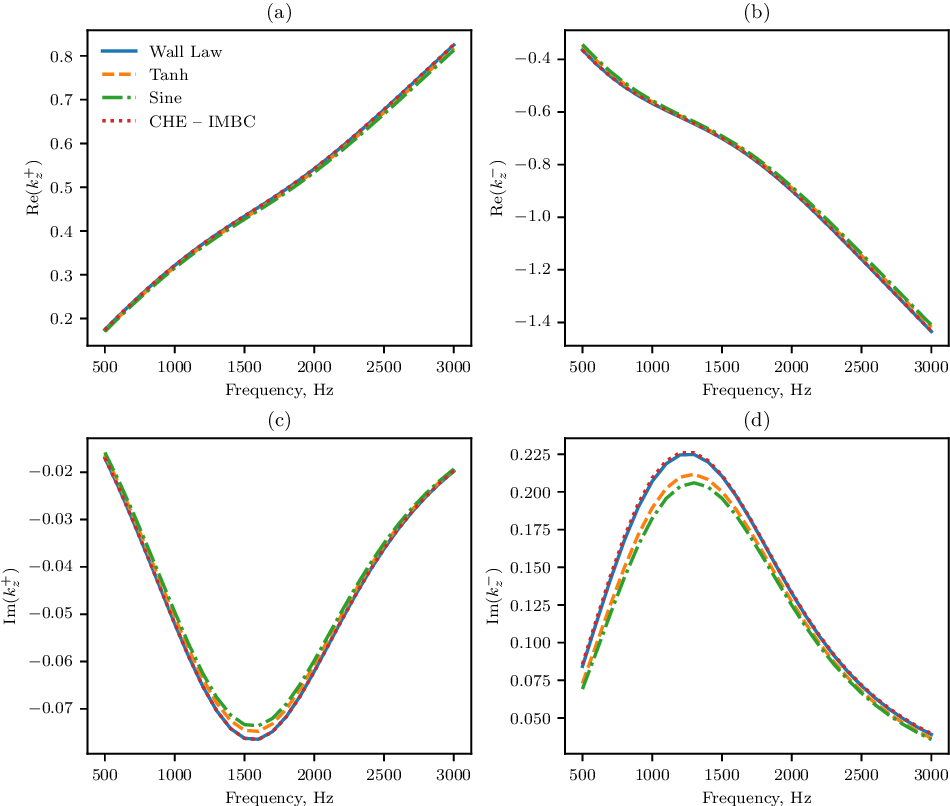}%
        \caption{Wavenumbers obtained for the SDOF-like impedance with different velocities distributions for the same $M$ and $\delta_{\SI{99.9}{\percent}}$.}%
        \label{fig:kz_SDOF_delta99}%
    \end{figure*}

    First, we examine the wavenumbers obtained from the Pridmore--Brown equation for different velocity profile shapes, all with the same boundary layer thickness, $\delta_{\SI{99}{\percent}}$, and average Mach number $M$. These results are compared with the wavenumbers derived from the Convected Helmholtz equation with the Ingard--Myers boundary condition, which models the physics within the boundary layer considering the average Mach number. For brevity, we focus initially on the case of $Z_{\text{SDOF}}$. The wavenumbers for the least attenuated mode, for both upstream ($k_z^-$) and downstream ($k_z^+$) propagation are presented in Figure~\ref{fig:kz_SDOF_delta99}. In this work, only the least attenuated mode is considered, as it is likely to dominate in the traditional impedance eduction range (i.e., low Helmholtz number and small ducts). However, this assumption may not hold for novel facilities designed to target higher-order mode eduction \mbox{\citep{yangShearFlowEffects2024, rashidi_3d_2025}}.
    Results suggest good agreement between all considered velocity profiles and the predictions from the CHE-IMBC for the real component of the wavenumber in both propagation directions. However, for the imaginary component of the wavenumber, which corresponds to the attenuation rate, significant differences are observed between the wavenumbers obtained for the hyperbolic tangent and sinusoidal flow profiles, compared to those obtained for the more realistic distribution given by the universal wall law, particularly for upstream propagation. On the other hand, good agreement is observed between the CHE-IMBC solution and the PBE solutions. This suggests that the IMBC may provide a good approximation for the typical range of frequencies and duct dimensions of impedance eduction. A possible explanation is that, due to the high gradient near the wall, even though the velocity distribution extends nearly across the entire duct half-width, the region where refractive effects are significant is confined to a much thinner region, making the infinitely thin hypothesis of the IMBC more appropriate than originally suggested.
    
\begin{figure*}%
        \centering%
        \includegraphics{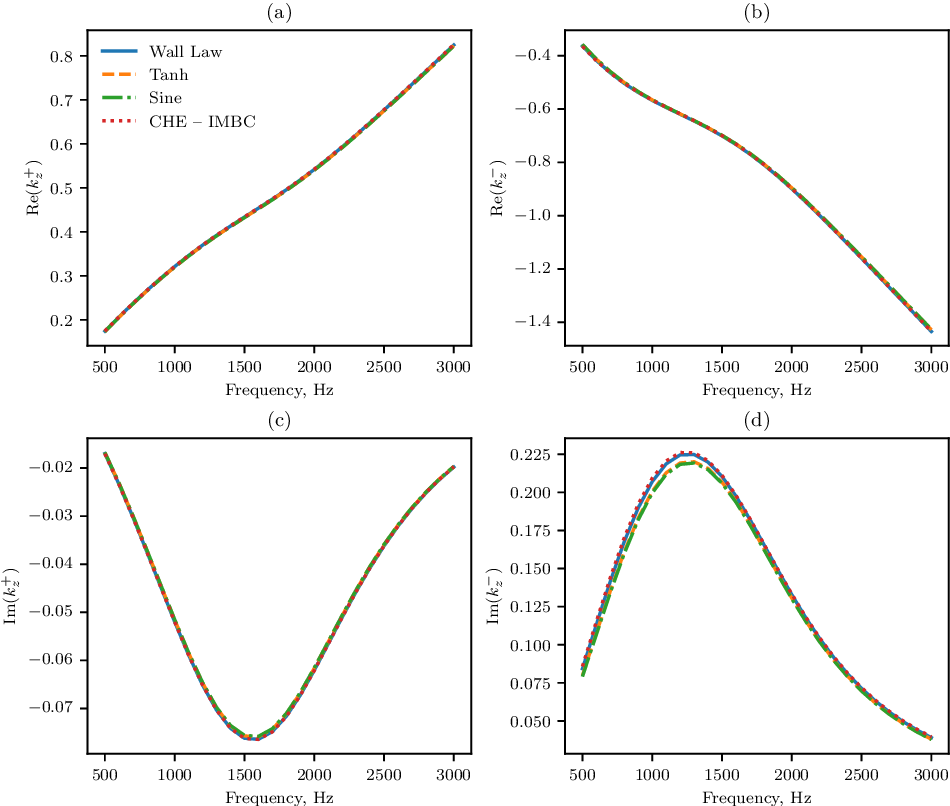}%
        \caption{Wavenumbers obtained for the SDOF-like impedance with different velocities distributions for the same $M$ and $\delta^*$.}%
        \label{fig:kz_SDOF_deltaStar}%
    \end{figure*}

\begin{figure*}%
        \centering%
        \includegraphics{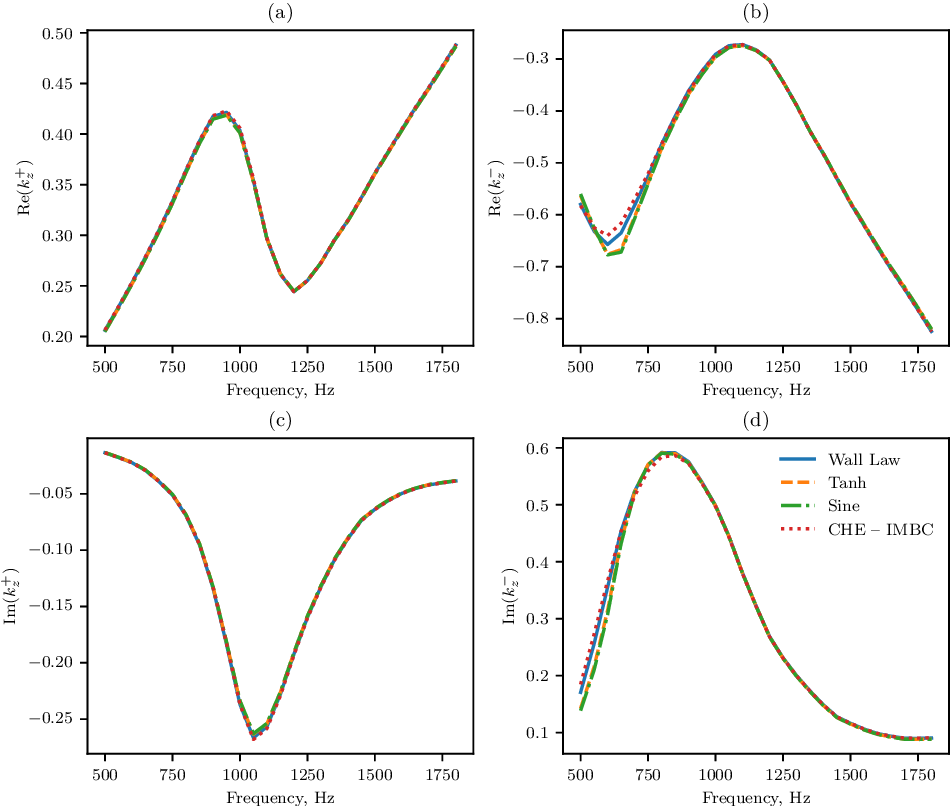}%
        \caption{Wavenumbers obtained for the digitized CT57 impedance with different velocities distributions for the same $M$ and $\delta^*$}%
        \label{fig:kz_CT57_deltaStar}%
    \end{figure*}

    Next, we consider the case where the different velocity distributions are adjusted to match the average Mach number and the boundary layer displacement thickness, $\delta^*$, rather than the boundary layer thickness, $\delta_{\SI{99}{\percent}}$. This approach is expected to improve the agreement between the acoustic attenuation predicted by the different flow profiles, particularly for downstream propagation~\citep{nayfeh1974Effect}. The wavenumbers for the least attenuated mode in both upstream and downstream propagation are shown in Figures~\ref{fig:kz_SDOF_deltaStar}
    and \ref{fig:kz_CT57_deltaStar},
    for the impedances $Z_{\text{SDOF}}$ and $Z_{\text{CT57}}$, respectively.    

    As expected, a better agreement is observed for the wavenumbers obtained for the different velocity distributions, especially for downstream propagation. However, for upstream propagation, the wavenumbers for the hyperbolic tangent and sinusoidal flow profiles agree well with each other, and subtle differences are observed compared to the solution for the wall law and the prediction for the CHE-IMBC. These initial results suggest that assuming a uniform flow and compensating for the refraction within the boundary layer using the Ingard--Myers boundary conditions provides better estimates of the acoustic field in impedance eduction facilities under typical test conditions, compared to solving for an explicit velocity distribution that is not representative of realistic conditions. Nevertheless, since the observed differences are subtle, it remains to be assessed whether they are significant in terms of impedance eduction, as will be investigated in the following.

\begin{figure*}%
        \centering%
        \includegraphics{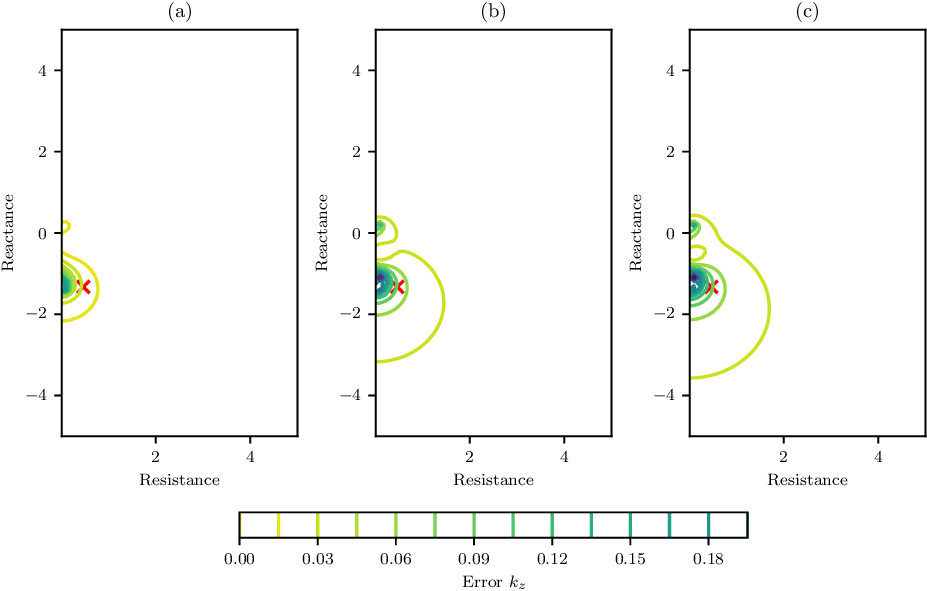}%
        \caption{Contour plots of the error of assuming the Ingard--Myers boundary condition as a simplification of different velocities profile formulations. Red "X" denote the value of the CT57 impedance at \SI{550}{\hertz}. (a) Universal Law of the Wall; (b) Hyperbolic Tangent and; (c) Sinusoidal.}%
        \label{fig:contour}%
    \end{figure*}

    It is worth noting that the difference between the wavenumbers obtained for the wall law velocity distribution and the estimation obtained with CHE-IMBC is notably higher at the lower frequency range for the impedance $Z_{\text{CT57}}$, as shown in Figure~\ref{fig:kz_CT57_deltaStar}d. To investigate this, we propose analysing the error between the estimation from the CHE-IMBC, $k_{z, \text{CHE}}$, and the exact solution from the PBE, $k_{z, \text{PBE}}$, defined as
    \begin{equation}
        \text{error} = \dfrac{|k_{z, \text{PBE}} - k_{z, \text{CHE}}|}{k_{z, \text{PBE}}},
    \end{equation}
    in the complex impedance plane. We fix the frequency at \SI{550}{\hertz} and consider the resistance range $\theta \in [0, 5]$ and the reactance range $\chi \in [-5, 5]$. The three flow profile formulations are considered, and the corresponding contour plots are shown in Figure~\ref{fig:contour}.

    Results suggest that the error function is almost zero for the majority of the considered impedance plane, with the notable exception of the region defined by resistances smaller than 1 and reactances between -2 and 0. This frequency range lies close to regions previously identified as being susceptible to surface-mode degeneracies and double roots~\citep{brambley2013surface}, which are known to compromise the stability and well-posedness of the Ingard--Myers boundary condition~\citep{brambley2009fundamental}. While a definitive modal crossing cannot be confirmed here, this mechanism provides a plausible explanation for the increased uncertainty observed.

\subsection{Effects of flow profiles on impedance eduction}

\begin{figure*}%
        \centering%
        \includegraphics{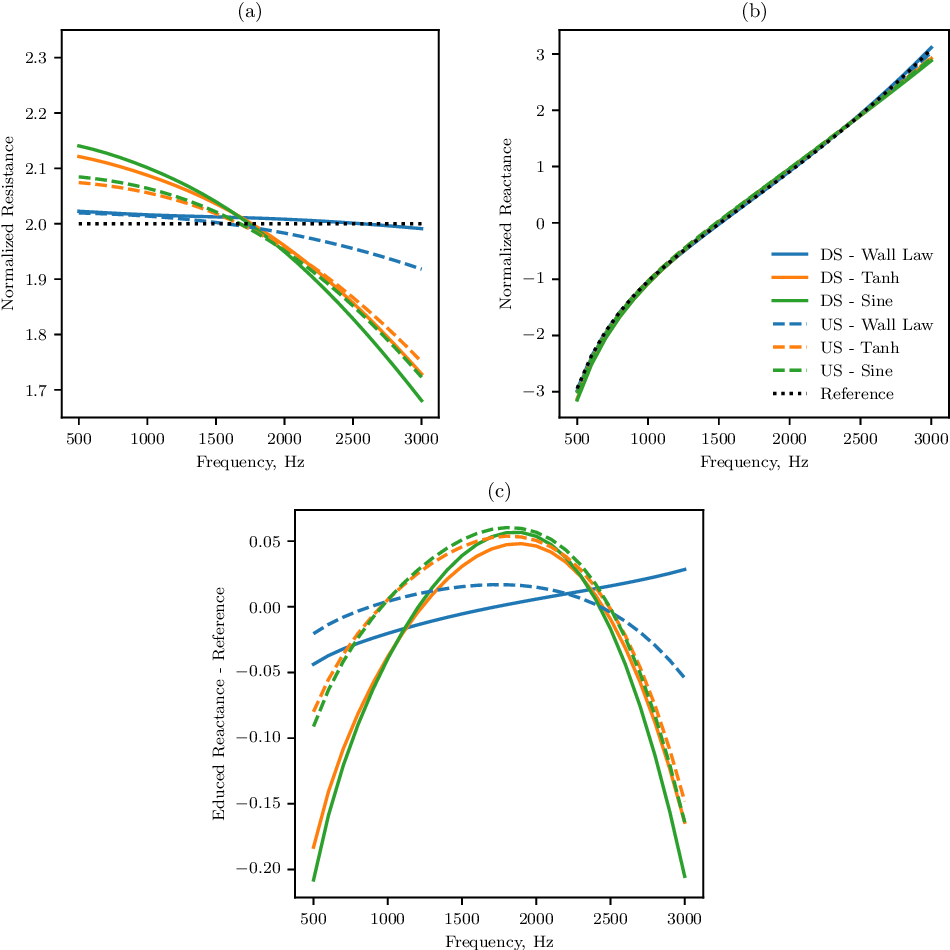}%
        \caption{Educed impedances obtained for the wavenumbers evaluated for the PBE considering different velocities profile shapes with the impedance $Z_{\text{SDOF}}$ at the lined wall. Velocities profiles match the universal wall law with $u_\tau = \SI{3.95}{\meter\per\second}$, in average Mach number and boundary layer displacement thickness $\delta^*$. US - Upstream Source (downstream propagation), and; DS - Downstream Source (upstream propagation).  (c) plots the difference between the reactances and the reference reactance that are plotted in (b).}%
        \label{fig:educedImpedances1}%
    \end{figure*}

    The next step, which is the main goal of this study, is to evaluate the impact of considering different flow velocity profiles on the evaluation of the acoustic field in impedance eduction. As discussed in Section~\ref{subsec:eduction}, we use the wavenumbers obtained for the least attenuated mode, considering the different velocity profiles, in the classical straightforward impedance eduction routine. This routine assumes uniform flow and the Ingard--Myers boundary condition to model the slip velocity at the wall. For the sake of brevity, from this point on, we will focus solely on the impedance $Z_{\text{SDOF}}$, as it is a more representative case of typical acoustic liners' impedance. The impedances educed with the proposed numerical experiment using Eqs.~\eqref{eq:eigenvalue} and \eqref{eq:dispRel}, along with the wavenumbers obtained for the different velocity profiles in the PBE, are shown in Figure~\ref{fig:educedImpedances1}.

    The impedances educed using the wavenumbers obtained from the solution of the hyperbolic tangent and sinusoidal flow profiles exhibit a similar trend regarding the mismatch observed experimentally between upstream and downstream acoustic sources (downstream and upstream propagation, respectively). At lower frequencies, the upstream source case results in a lower resistance, with the opposite trend observed at higher frequencies. This behaviour is similar to what has been observed by \citet{roncen2020WavenumberBased}; however, in our case, the reference impedance is not the midpoint between the two curves. For the most realistic flow profile, the wall law, the conclusions differ significantly. At the lower frequency end, the assumption of uniform flow with the IMBC introduces a small bias for both acoustic source positions, with good agreement observed between them. At higher frequencies, the curves diverge, with the upstream source (downstream propagation) surprisingly showing a greater deviation from the reference impedance.

    The trends observed in Figure~\ref{fig:educedImpedances1} indicate a clear frequency-dependent behaviour of the impedance eduction. At lower frequencies, the resistance tends to be overpredicted for all flow-profile assumptions, whereas at higher frequencies it is systematically underpredicted. This behaviour appears largely independent of the specific velocity profile and suggests that it is primarily associated with the limitations of the uniform-flow + IMBC approximation rather than with the detailed structure of the boundary layer.
    
    Differences between flow-profile assumptions become more pronounced at higher frequencies, where the educed impedance shows increased sensitivity to the near-wall velocity gradient. In this regime, simplified sheared profiles lead to larger deviations, while the wall-law profile yields results closer to the reference case. The larger upstream–downstream discrepancy observed for the wall-law case and the change in trend around 1.6–\SI{1.7}{kHz} are consistent with a transition toward increased modal sensitivity in the duct.

    The results obtained so far in this work suggest that the shape of the velocity profile considered when solving for the acoustic field in small ducts with lined walls plays a significant role. Additionally, the Ingard–Myers boundary condition provides better estimations, when compared to the exact solution for simplified velocity distributions. However, a single duct geometry and a single average Mach number have been considered. In what follows, we propose a parametric study on the duct geometry and bulk Mach number impact on our conclusions.

\subsection{Parametric study}

    In this section, we analyse the sensitivity of the IMBC accuracy to variations in the average Mach number and duct width through a parametric analysis. This is particularly important given that recent impedance eduction facilities are moving toward higher bulk Mach numbers and larger duct cross-sections, often with a specific focus on multimodal acoustic propagation (e.g.\ \cite{yangShearFlowEffects2024, rashidi_3d_2025}). We consider the wavenumbers obtained by solving the PBE and the flow profile described by the universal law of the wall.
    Furthermore, the thickness of the boundary layer may influence the accuracy of the Ingard–Myers boundary condition. However, since it has been observed that the universal law of the wall formulation used in this work can extend to the entire half-width of the duct, the boundary layer thickness in this case is a function of both the duct width and the viscosity. To produce significant variations in $\delta$ by changing the viscosity $\nu$, non-realistic values would need to be considered. For this reason, we have decided not to include the boundary layer thickness as a parameter in this parametric analysis.
    
\begin{figure*}%
        \centering%
        \includegraphics{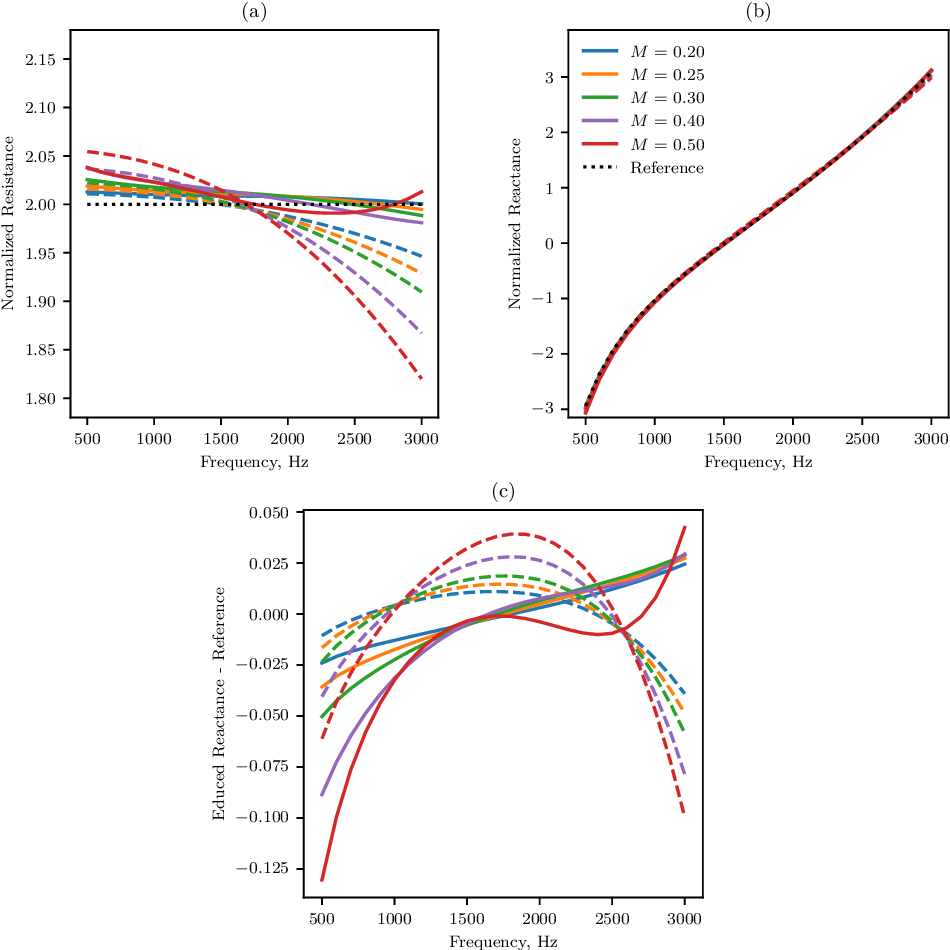}%
        \caption{Parametric study on the effect of the average Mach number. Impedances educed with the IMBC for the wavenumbers obtained for the exact solution of the PBE with a realistic flow profile. Solid lines: downstream acoustic source; dashed lines: upstream acoustic source.  (c) plots the difference between the reactances and the reference reactance that are plotted in (b).}%
        \label{fig:parametricMach}%
    \end{figure*}

    First, we examine the effect of the average Mach number. The duct width is set to $W = \SI{40}{\milli\meter}$, and the air viscosity is $\nu = \SI{1.48}{\meter^2\per\second}$. The friction velocity was adjusted to vary with the average Mach number, and the values considered are presented in Table~\ref{tab:uTauM}.
    \begin{table}[t]%
        \centering%
        \caption{Friction velocities $u_\tau$ considered for the parametric analysis as a function of the average Mach number $M$ for a duct width of $W=\SI{40}{\milli\meter}$.}%
        \label{tab:uTauM}%
        \vspace{1ex}%
        \begin{tabular}{r|lllll}
         $M$                               & 0.20 & 0.25 & 0.30 & 0.40 & 0.50 \\\hline
         $u_\tau$ [\si{\meter\per\second}] & 2.93 & 3.58 & 4.23 & 5.49 & 6.73
        \end{tabular}%
    \end{table}%
    The impedances educed for the different average velocities are shown in Figure~\ref{fig:parametricMach}.

    The results indicate that the deviation in the educed impedance obtained using a uniform-flow model combined with the Ingard–Myers boundary condition increases with the bulk Mach number, particularly for the resistance. Here, the impedance obtained using the PBE with a realistic turbulent flow profile is treated as the reference solution. This aligns with the observations from the evaluation of the wavenumbers, where the largest differences are noted in the imaginary component of the wavenumber. As with the resistance, the imaginary component is related to acoustic dissipation. These findings are consistent with previous studies reported in the literature~\citep{yangShearFlowEffects2024}.
    
\begin{figure*}%
    \centering%
    \includegraphics{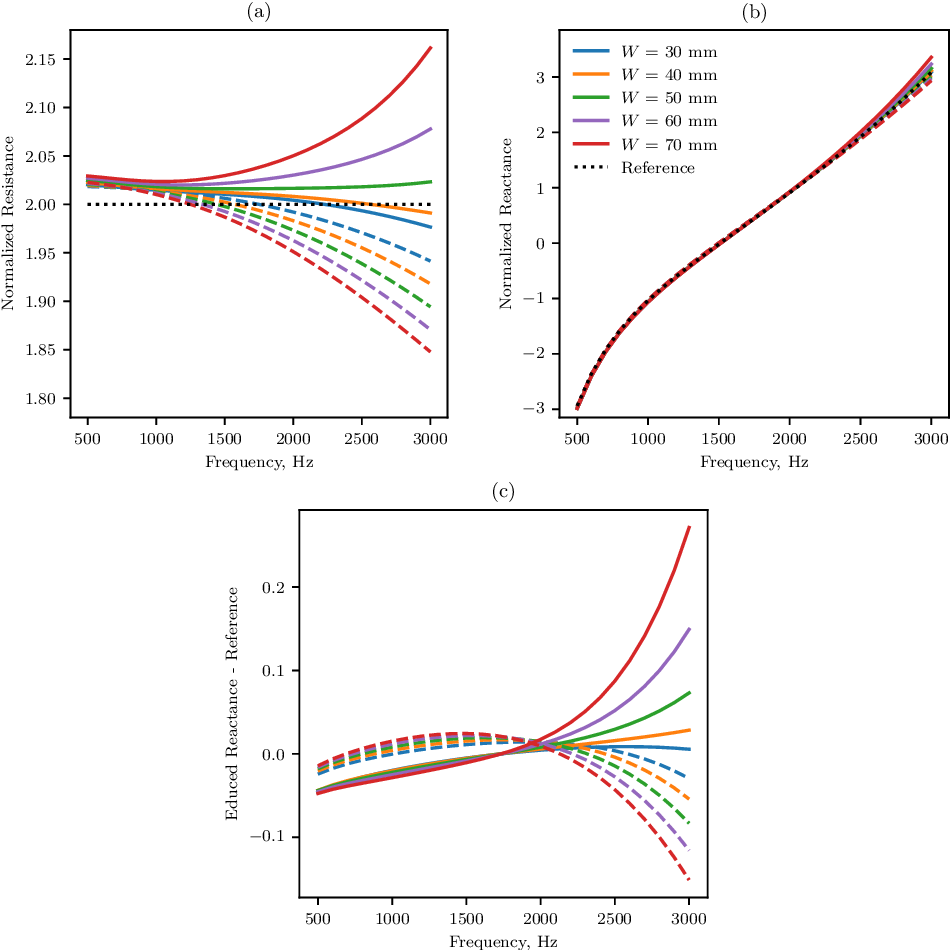}%
    \caption{Parametric study on the effect of the duct width. Impedances educed with the IMBC for the wavenumbers obtained for the exact solution of the PBE with a realistic flow profile. Solid lines: downstream acoustic source; dashed lines: upstream acoustic source.  (c) plots the difference between the reactances and the reference reactance that are plotted in (b).}%
    \label{fig:parametricWidth}%
\end{figure*}

    Next, we examine the effect of the duct width—and consequently, the boundary layer thickness—on the accuracy of the IMBC for impedance eduction. This analysis considers the typical dimensions of traditional liner impedance eduction facilities, which typically feature duct widths smaller than \SI{70}{\milli\meter}. Novel approaches, such as curved~\citep{carr2024acoustic} and multimodal~\citep{humbert2022multimodal} duct configurations, are not considered in this study. The average Mach number is fixed at the same value as the baseline case, $M = 0.279$, and the friction velocity is adjusted for each considered duct width. The duct width values and corresponding friction velocities are summarised in Table~\ref{tab:uTauW},
    \begin{table}%
        \centering%
        \caption{Friction velocities $u_\tau$ considered for the parametric analysis as a function of the duct width $W$ for an average Mach number of 0.279.}%
        \label{tab:uTauW}%
        \vspace{1ex}%
        \begin{tabular}{r|lllll}
         $W$ [\si{\milli\meter}] & 30 & 40 & 50 & 60 & 70 \\\hline
         $u_\tau$ [\si{\meter\per\second}] & 4.06 & 3.95 & 3.87 & 3.81 & 3.76
        \end{tabular}%
    \end{table}%
    while the educed impedances for the different cases are presented in Figure~\ref{fig:parametricWidth}.

    The results suggest that the accuracy of the IMBC decreases with increasing duct width, particularly in predicting the resistive component of the impedance. For the reactance, larger errors are observed at the higher frequency range, especially for the downstream acoustic source (upstream propagation). These findings align with expectations, as an increase in duct width leads to a corresponding increase in the dimensional boundary layer thickness. This deviates further from the infinitely thin boundary layer assumption of the IMBC.

    \subsection{Summary of theoretical results} \label{sec:partConclu}

        Summarizing the numerical results presented in the preceding sections:
        \begin{itemize}
            \item Matching the bulk Mach number and boundary-layer displacement thickness is not sufficient to ensure the same acoustic field across different velocity-profile shapes;
            \item For typical impedance eduction test facilities, a better approximation to the acoustic field in a realistic turbulent flow profile is provided by assuming uniform flow combined with the Ingard–Myers boundary condition than assuming a sheared flow with a simplified flow profile, and;
            \item The error associated with the uniform-flow + IMBC assumption tends to increase with increasing bulk Mach number and boundary-layer thickness and/or duct height.
        \end{itemize}

\section{Application to Experimental Data} \label{sec:experiments}

    Finally, we propose to extend the analysis of this study to experimental data gathered at the Liner Impedance Test Rig of the Federal University of Santa Catarina (LITR/UFSC). The test rig's test section consists of modular rectangular cross-sectioned ducts measuring $\num{40}\times\SI{100}{\milli\meter^2}$ (i.e., $W = \SI{40}{\milli\meter}$). Quasi-anechoic terminations at the test rig inlet and outlet minimise acoustic reflections. Eight Beyma CP-855nD compression drivers are distributed both upstream and downstream of the liner test sample holder to generate sound fields up to \SI{150}{\deci\bel}, propagating either with or against the flow towards the liner sample.
    
    An external compressed air system provides the flow supply, capable of sustaining a cross-section averaged flow up to Mach \num{0.7}. A Pitot tube located at the test rig inlet is used to control and monitor the flow Mach number during tests. The average Mach number in the lined section is derived from the Pitot tube measurement using a pre-calibrated factor determined through a quadrature method, as defined by the standard ISO~3966:2008\mbox{\citep{ISO3966}}, with a 5 by 5-points grid. The liner sample holder has an opening for liner samples with a maximum length of \SI{420}{\milli\meter}.
    
    An array of sixteen equally spaced flush-mounted B\&K DeltaTron 4944 1/4" pressure field microphones is installed on the wall opposite the liner section for impedance eduction. The spacing between consecutive microphones is \SI{20}{\milli\meter}. In this work, half of the microphones are skipped, resulting in an effective separation of \SI{40}{\milli\meter} to reduce uncertainties in the lower attenuation range of the liner~\citep{bonomo2020Parametric}. Signals are recorded using a National Instruments PXIe-4499 data acquisition (DAQ) module at a sampling rate of \SI{25.6}{\kilo\hertz}. Measurements are conducted using a harmonic excitation signal, which also serves as a reference for cross-spectrum estimation using Welch's method, with \num{30} averages of \num{25600} samples and \SI{75}{\percent} overlap. All hardware control, signal processing, and data post-processing are performed using in-house Python3 code.
    
    Two liner samples are employed in this study, referred to as samples A and B. Both samples are typical single-degree-of-freedom liner constructions, each with a length of \SI{420}{\milli\meter}. A summary of the relevant parameters for both samples is presented in Table~\ref{tab:liners}. For both samples, no significant non-linear effects due to the SPL are expected.
    \begin{table}%
        \centering%
        \caption{Liner samples parameters. $\sigma$ - percentage of open area; $h$ - cavity height; $d$ - holes diameter, and; $t$ - perforate sheet thickness.}%
        \label{tab:liners}%
        \vspace{1ex}%
        \begin{tabular}{lcccc}
            \hline
            \multicolumn{1}{l|}{Parameter} & $\sigma$ [\si{\percent}] & $h$ [\si{mm}] & $d$ [\si{mm}] & $t$ [\si{mm}] \\ \hline
            \multicolumn{1}{l|}{Sample A} & 5 & 40 & 1.2 & 1 \\
            \multicolumn{1}{l|}{Sample B} & 12 & 25.4 & 0.835 & 1 \\ \hline
        \end{tabular}%
    \end{table}%

    Tests were conducted under three different flow conditions: in the absence of flow ($M=0$); and with bulk Mach numbers (area-averaged) of $M = 0.2$ and $M = 0.3$. A stepped pure-tone excitation was employed in a frequency range from \SI{500}{\hertz} to \SI{3000}{\hertz}, with increments of \SI{100}{\hertz}. The sound pressure level was set to \SI{130}{\decibel} for the plane wave amplitude propagating towards the liner, with the acoustic source positioned either upstream or downstream (one at a time) of the liner.  
    
    We consider four cases for impedance eduction using the experimental acoustic field. First, we examine the traditional straightforward method, which assumes that acoustic propagation is governed by the CHE and that the Ingard--Myers boundary condition applies to the lined walls. The other three cases involve solving the PBE while considering different flow velocity distributions. In the first of these, we assume that the flow profile can be approximated by the universal law of the wall, Eq.~\eqref{eq:vanDriest}. The friction velocity is adjusted so that the average Mach number of the 1D profile matches the bulk Mach number of the 2D test section. This approach follows the conclusion of \citet{jingInvestigationStraightforwardImpedance2015}, who demonstrated that, when simplifying a 3D duct to a 2D duct, the average Mach number must remain constant.  
    
    The other two cases also solve the PBE but use a hyperbolic tangent velocity distribution. In the first of these, the boundary layer thickness $\delta_{\SI{99}{\percent}}$ is matched to that of the universal law of the wall. In the final case, the boundary layer displacement thickness, $\delta^*$, is matched instead.

    For the impedance eduction considering the solution of the PBE, we follow an iterative procedure first presented by \citet{roncen2020WavenumberBased}. The dominant axial wavenumber of sound in the lined duct section, $k_{z, \text{exp}}$, is extracted from the equally spaced microphone array record using the KT algorithm, as detailed in \citet{bonomo2020Parametric}. The eduction routine minimises a cost function defined as  
    \begin{equation}
        \mathcal{F}(Z) = \left| k_{z, \text{exp}} - k_{z, \text{PBE}}(Z) \right|,
    \end{equation}
    where $k_{z, \text{PBE}}$ is the wavenumber obtained by solving the eigenvalue problem of the PBE. The Levenberg--Marquadt algorithm \mbox{\citep{levenberg-1944,marquardt-1963}} is used to minimize this cost function. To accelerate convergence, the impedance obtained by solving the convected Helmholtz equation with the Ingard--Myers boundary condition is used as an initial guess.  
    
\begin{figure*}%
        \centering%
        \includegraphics{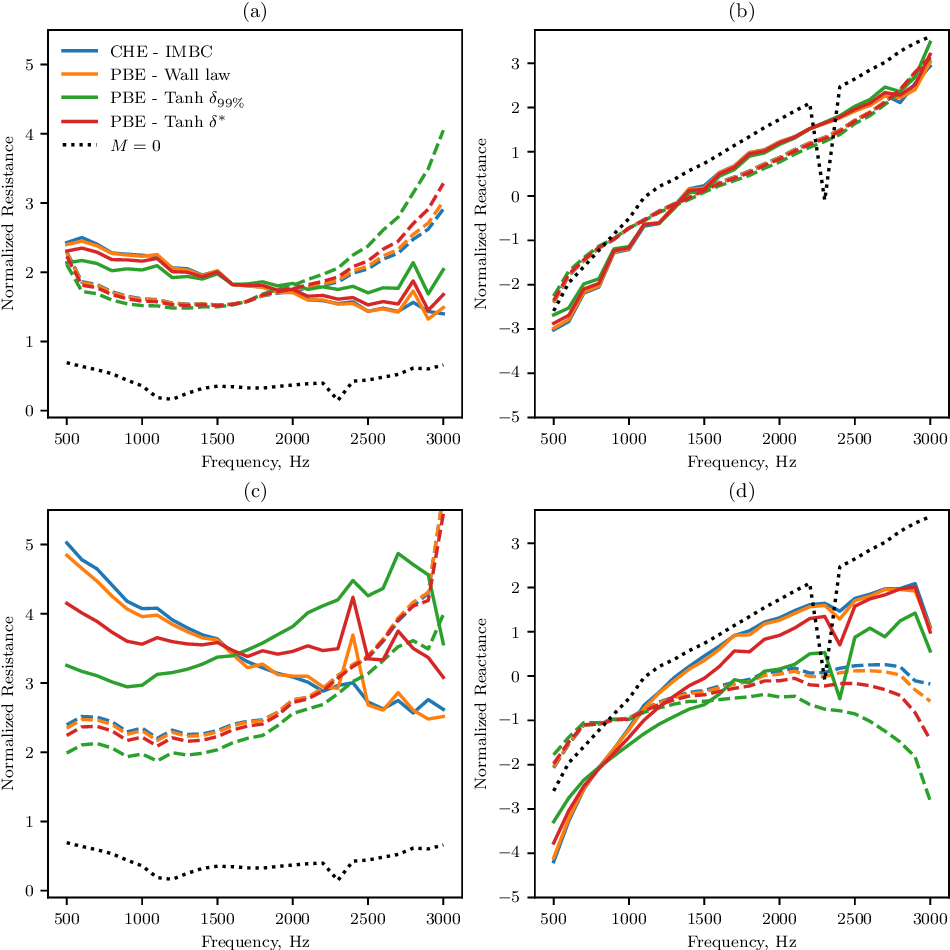}%
        \caption{Impedances educed for sample A. (a,b) $M = 0.2$; (c,d) $M = 0.3$. Dashed lines denote impedances educed with an upstream source, and solid lines denote impedances educed with a downstream source.}%
        \label{fig:barril}%
    \end{figure*}

\begin{figure*}%
        \centering%
        \includegraphics{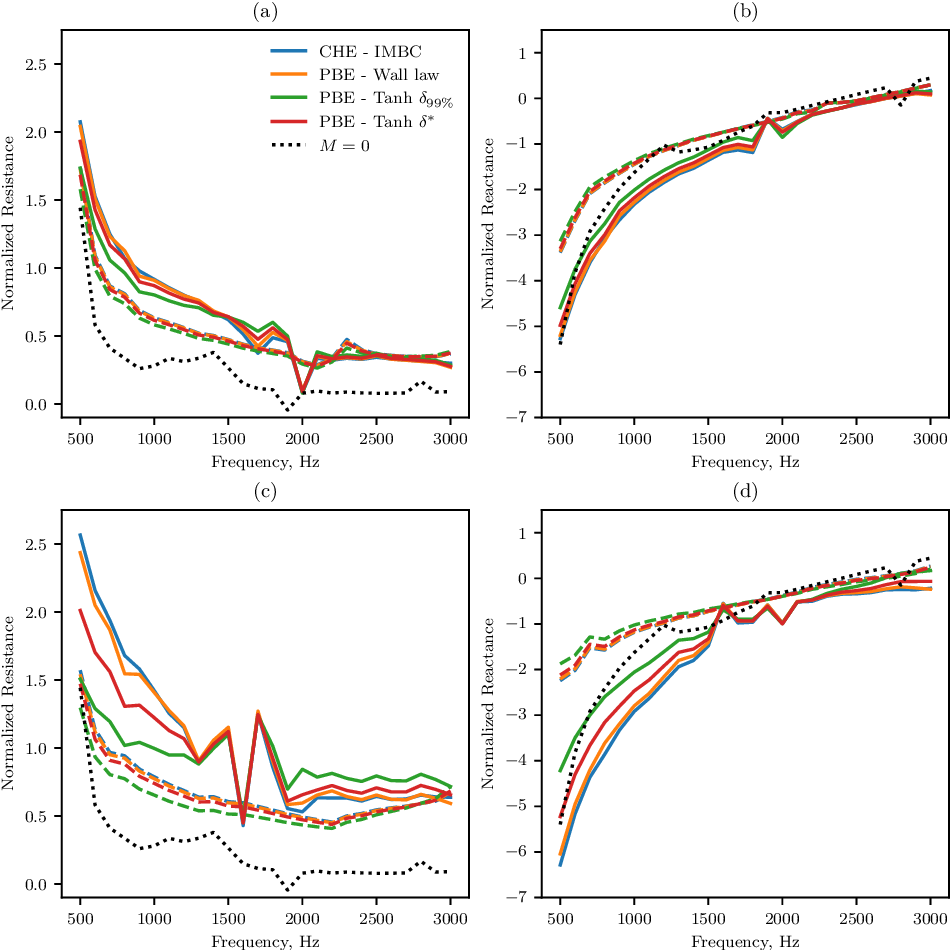}%
        \caption{Impedances educed for sample B. (a,b) $M = 0.2$; (c,d) $M = 0.3$. Dashed lines denote impedances educed with an upstream source, and solid lines denote impedances educed with a downstream source.}%
        \label{fig:utas}%
    \end{figure*}

    The results obtained for the four cases considered, using samples A and B, are shown in Figures~\ref{fig:barril}
    and \ref{fig:utas},
    respectively. Greater differences between upstream and downstream educed impedances are observed with sample A, which exhibits stronger non-linear behaviour with respect to flow effects. However, the conclusions regarding the impact of assuming different 1D flow velocity distributions are consistent for both liners and can be summarised as follows. The differences in the educed resistances for the different velocity profiles are larger compared to the differences in the reactances, aligning with observations from the numerical experiment. Greater differences are also noted with increasing Mach number. One may notice that around 1400 Hz, the duct cross-section allows more than one propagating mode, which increases the sensitivity of the impedance eduction process and may contribute to the local slope change observed in the reactance, particularly in the absence of mean flow.

    Additionally, the impact of assuming different flow profiles is more pronounced for a downstream acoustic source (upstream propagation), consistent with the larger biases observed in the wavenumbers obtained for the different formulations. Regarding the different formulations used for solving the PBE, significant differences are observed among the three cases. The case with matching $\delta^*$ shows the best agreement between the solution obtained for the hyperbolic tangent profile and that for the universal law of the wall.  
    
    As anticipated by the numerical experiment, good agreement is observed between the prediction using the Ingard--Myers boundary condition and the solution obtained for the universal law of the wall profile, particularly for an upstream acoustic source (downstream propagation). For impedances educed with a downstream acoustic source, the IMBC slightly over-predicts compared with the solution of the PBE with the law of the wall. This is consistent with the results of \citet{weng2018Flow}, who solved the linearised Navier--Stokes equation for a realistic flow profile comparable to the one considered in the present work. In summary, these results indicate that, in the low-Mach-number and small-duct regime considered here, neglecting the velocity gradient altogether by assuming uniform flow leads to more accurate impedance eduction results than adopting overly simplified sheared velocity profiles, whose artificial refraction effects can bias the solution.  Note that this is in agreement with the theoretical results summarized in section~\ref{sec:partConclu}.

\section{Conclusion} \label{sec:conclu}

    In this work, the effects of the sheared flow profile shape on acoustic propagation in a 2D duct were revisited, extending the early work of \citet{nayfeh1974Effect} to the context of impedance eduction techniques. Three velocity profiles were considered for solving the acoustic field in a duct with sheared grazing flow, using the Pridmore--Brown equation, with the profiles matching the average Mach number and either the $\delta_{\SI{99}{\percent}}$ or $\delta^*$ boundary layer thicknesses. The wavenumbers obtained from the PBE were then compared to those estimated by solving the case with uniform flow, i.e., the Convected Helmholtz equation, with the Ingard--Myers boundary condition handling the refractive effects within the boundary layer. Results suggest that the IMBC leads to lower errors relative to the solution for a sheared flow profile with a realistic wall law turbulent boundary layer velocity distribution, compared to the solution obtained with a simplified profile formulation. Consistent with the findings of \citet{nayfeh1974Effect}, it was observed that matching the boundary layer displacement thickness $\delta^*$ improves agreement between the different formulations, although noticeable differences remain.  

    Next, a numerical experiment was conducted to assess the accuracy of the IMBC in the typical impedance eduction range for small ducts, by assuming that the PBE is a exact representation of the real world. The wavenumbers obtained for the different flow profile formulations were used as input for the traditional straightforward impedance eduction routine. Results suggest that for non-realistic flow profile formulations, the simplification to the uniform flow assumption with the IMBC may lead to mismatches between results obtained for upstream and downstream propagating waves, particularly for the acoustic resistance. However, this mismatch does not occur when a realistic velocity distribution is used to simulate real-world conditions. In contrast, several previous studies have concluded that the Myers condition can be inaccurate, not only because of its ill-posedness in the time domain but also in predicting absorption in the frequency domain. The present findings go against this prevailing view.  In fact, most of these earlier assessments have relied on simplistic velocity profiles less representative of real flows than the van Driest profile, whereas the present work demonstrates improved agreement when a realistic profile is used.  In particular, the van Driest profile has a far higher velocity gradient at the lined wall than the more simplistic velocity profiles, and so might be expected to be better approximated by the vortex sheet assumption of the IMBC than the smoother more simplistic velocity profiles.  This can also be noted in the velocity profile shape factors~\eqref{equ:shape-factor}, with a factor of two difference between the van Driest and other profiles. In addition, a parametric study was conducted to investigate the impact of the average Mach number and the duct width, and consequently, the boundary layer thickness. It was found that the error associated with the simplification to the IMBC increases with both the average Mach number and the duct width.  
    
    Hereinafter, an iterative eduction routine was used to evaluate the impact of solving for the sheared flow profile rather than relying on the IMBC approximation with experimental acoustic data. This analysis allowed to investigate the effect of simplifying the flow profile representation when performing impedance eduction and solving for the sheared flow. The results obtained align with the conclusions of \citet{weng2018Flow}, who suggested that solving for the acoustic field using a realistic flow profile produces reasonable agreement with the IMBC solution.  
    
    The main conclusion of this work is that the Ingard--Myers boundary condition is a reasonable simplification in the context of low Mach number and small-duct impedance eduction, at least for 2D ducts. These findings contrast with the interpretation of \citet{gabard2013Comparison}, who noted no significant influence of the boundary-layer profile on acoustic propagation provided the correct boundary layer thickness was maintained. The present results demonstrate that, in the specific context of impedance eduction in small ducts, the shape of the velocity profile can indeed have a measurable and physically meaningful impact on the educed impedance, independently from the boundary-layer thickness. The natural continuation of this work is its extension to a realistic 3D duct, as proposed by \citet{roncen2020WavenumberBased}, while taking into account the importance of a realistic flow profile representation. Finally, it is important to highlight that the current conclusions are limited to the specific case of small ducts with only plane waves, and it remains to be seen how they extend to larger ducts, higher frequencies, and higher-order modes.  In particular, higher-order modes, and especially the higher-azimuthal-order modes in cylindrical ducts typical of rotor-alone tones in aeroengines, experience a strong refractive effect from the boundary layer that could be expected to differ significantly from the plane-wave modes considered here.  However, lab-based impedance eduction tests of acoustic liners usually also rely on small-ducts with plane waves, and so the neglect of higher-order modes in this work remains valid in that context.

\section*{Data Availability}
    Data will be made available upon request.

\section*{Conflict of Interest}
    The authors declare no conflict of interest.

\section*{Acknowledgments}
    On behalf of  L.A.~Bonomo and J.A.~Cordioli, this research was partially funded by CNPq (National Council for Scientific and Technological Development).
    L.A.~Bonomo acknowledges that this study was financed in part by the Coordenação de Aperfeiçoamento de Pessoal de Nível Superior – Brasil (CAPES), Finance Code 001.
    E.J~ Brambley gratefully acknowledges the support of the UK Engineering and Physical Sciences Research Council (EPSRC grant EP/V002929/1).

\appendix
\section{Assessment of the Convergence of the Numerical Grid} \label{ap:convergence}

\begin{figure*}%
        \centering%
        \begin{subfigure}{0.495\linewidth}%
            \includegraphics{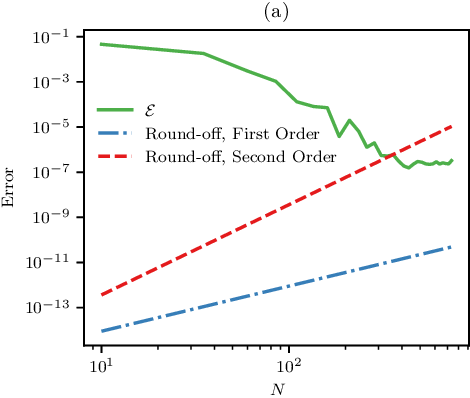}%
        \end{subfigure}%
        \begin{subfigure}{0.495\linewidth}%
            \includegraphics{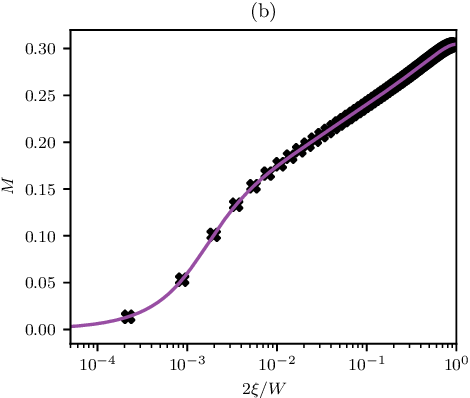}%
        \end{subfigure}%
        \caption{(a) Convergence error of the least-attenuated mode wavenumber propagating upstream at \mbox{\SI{3}{\kHz}} with the SDOF-like impedance. (b) Visualisation of the Chebyshev grid points over the universal law of the wall profile for $N = 151$.}%
         \label{fig:convergence}%
    \end{figure*}

    The round-off error in the $i$-th derivative based on Chebyshev polynomials is a function of the machine precision, $\epsilon$, and the number of points on the grid, $N$. The maximum value that this error can achieve is given by \mbox{\citep{don1997Accuracy}}
    \begin{equation}
        \text{round-off error} = \epsilon \left( \dfrac{2 N}{\pi} \right)^{2i}.
    \end{equation}
    This implies that the grid cannot be refined indiscriminately, since the eigenpairs will not converge to the exact solution. On the other hand, it must be sufficiently refined to accurately represent the boundary layer profile.
    
    To evaluate whether the grid considered in this work converges to the exact solution, we estimate the convergence error, $\mathcal{E}$, as
    \begin{equation}
        \mathcal{E} \equiv \left| \dfrac{k_{z,N} - k_{z,N_{\text{max}}}}{k_{z,N_{\text{max}}}} \right|,
    \end{equation}
    where $k_{z,N}$ is the wavenumber obtained for a grid with $N$ points and $N_{\text{max}}$ is the maximum number of points considered in the grid. For this work, $N_{\text{max}} = 785$ was found to be sufficiently large. The convergence obtained for the upstream propagating mode at the frequency of \mbox{\SI{3}{\kHz}} with the SDOF-like impedance is shown in Figure~\mbox{\ref{fig:convergence}}a; this analysis is typical of all frequencies and results computed here, and further convergence figures for other cases are omitted for brevity.
    
    The rate of convergence is observed to be approximately geometric up to $N \approx 300$. Beyond this point, the results suggest that convergence has been achieved. However, one may also notice that for $N > 200$, the round-off error associated with second-order derivatives begins to dominate. To achieve a balance between computational cost and accuracy, $N = 151$ was selected, which yields $\mathcal{E} < \mbox{\num{1e-6}}$. Finally, it is important to highlight that this discretization ensures at least one point in the linear viscous sublayer ($y^+\leq5$), as can be see in Figure~\mbox{\ref{fig:convergence}}b.

\raggedright
\bibliographystyle{elsarticle-num-names} 
\bibliography{references.bib}

@string(JFM="J.~Fluid Mech.")

@string(JSV="J.~Sound Vib.")

@string(JASA="J.~Acoust. Soc. Am.")

@string(AIAAJ="{AIAA}~J.")

@string(IJA=" Int. J. Aeroacoust.")

@article{brambley2013surface,
  title={Surface modes in sheared boundary layers over impedance linings},
  author={Brambley, E J},
  journal=JSV,
  volume={332},
  number={16},
  pages={3750--3767},
  year={2013},
  publisher={Elsevier},
  doi={10.1016/j.jsv.2013.02.028}
}

@article{van1956turbulent,
  title={On turbulent flow near a wall},
  author={van Driest, Edward R},
  journal={J.~Aeronaut. Sci.},
  volume={23},
  number={11},
  pages={1007--1011},
  year={1956},
  doi={10.2514/8.3713}
}

@inproceedings{rienstra2008spatial,
  title={Spatial instability of boundary layer along impedance wall},
  author={Rienstra, Sjoerd and Vilenski, Gregory},
  booktitle={14th AIAA/CEAS Aeroacoustics Conference},
  pages={2932},
  year={2008},
  doi={10.2514/6.2008-2932}
}

@inproceedings{carr2024acoustic,
  title={Acoustic Mode Decomposition in Rectangular Ducts With Sheared Flow},
  author={Carr, Alexander N},
  booktitle={30th AIAA/CEAS Aeroacoustics Conference},
  pages={3368},
  year={2024},
  doi={10.2514/6.2024-3368}
}

@inproceedings{humbert2022multimodal,
  title={Multimodal characterisation of acoustic liners using the MAINE Flow facility},
  author={Humbert, Thomas and Golliard, Joachim and Portier, Eric and Gabard, Gwenael and Auregan, Yves},
  booktitle={28th AIAA/CEAS Aeroacoustics Conference},
  pages={3082},
  year={2022},
  doi={10.2514/6.2022-3082}
}

@article{auregan2018stress,
  title = {On the Use of a Stress\textendash Impedance Model to Describe Sound Propagation in a Lined Duct with Grazing Flow},
  author = {Aur{\'e}gan, Yves},
  year = {2018},
  month = may,
  journal = JASA,
  volume = {143},
  number = {5},
  pages = {2975--2979},
  publisher = {{Acoustical Society of America}},
  issn = {0001-4966},
  doi = {10.1121/1.5037585}
}

@inproceedings{boden2016Effect,
  title = {On the Effect of Flow Direction on Impedance Eduction Results},
  booktitle = {22nd {AIAA/CEAS} Aeroacoustics Conference},
  author = {Bod{\'e}n, Hans and Zhou, Lin and Cordioli, Julio A. and Medeiros, Augusto A. and Spillere, Andr{\'e} M.N.},
  year = {2016},
  doi = {10.2514/6.2016-2727},
  pages = {2727},
}

@article{bonomo2020Parametric,
  title = {Parametric {{Uncertainty Analysis}} for {{Impedance Eduction Based}} on {{Prony}}'s {{Method}}},
  author = {Bonomo, Lucas A. and Spillere, Andr{\'e} M.N. and Cordioli, Julio A.},
  year = {2020},
  month = apr,
  journal = AIAAJ,
  volume = {58},
  number = {8},
  pages = {3625--3638},
  publisher = {{American Institute of Aeronautics and Astronautics}},
  issn = {00011452},
  doi = {10.2514/1.J059071}
}

@book{boyd2001Chebyshev,
  title = {Chebyshev and {{Fourier Spectral Methods}}},
  author = {Boyd, J.P.},
  year = {2001},
  edition = {2nd},
  publisher = {Dover},
  isbn = {9780486411835}
}

@article{brambley2009fundamental,
  title = {Fundamental Problems with the Model of Uniform Flow over Acoustic Linings},
  author = {Brambley, Edward James},
  year = {2009},
  journal = JSV,
  volume = {322},
  number = {4-5},
  pages = {1026--1037},
  issn = {0022460X},
  doi = {10.1016/j.jsv.2008.11.021}
}

@article{dean1974insitu,
  title = {An in Situ Method of Wall Acoustic Impedance Measurement in Flow Ducts},
  author = {Dean, P.D.},
  year = {1974},
  month = may,
  journal = JSV,
  volume = {34},
  number = {1},
  pages = {97--IN6},
  publisher = {{Academic Press}},
  issn = {0022-460X},
  doi = {10.1016/S0022-460X(74)80357-3}
}

@article{elnady2009validation,
  title = {Validation of an Inverse Semi-Analytical Technique to Educe Liner Impedance},
  author = {Elnady, T. and Bod{\'e}n, H. and Elhadidi, B.},
  year = {2009},
  journal = AIAAJ,
  volume = {47},
  number = {12},
  pages = {2836--2844},
  issn = {0001-1452},
  doi = {10.2514/1.41647}
}

@article{ferrante2016back,
  title = {Back-to-Back Comparison of Impedance Measurement Techniques Applied to the Characterization of Aero-Engine Nacelle Acoustic Liners},
  author = {Ferrante, Piergiorgio and De Roeck, Wim and Desmet, Wim and Magnino, Nicola},
  year = {2016},
  journal = {Applied Acoustics},
  volume = {105},
  pages = {129--142},
  issn = {1872910X},
  doi = {10.1016/j.apacoust.2015.12.004}
}

@article{gabard2013Comparison,
  title = {A Comparison of Impedance Boundary Conditions for Flow Acoustics},
  author = {Gabard, Gw{\'e}na{\"e}l},
  year = {2013},
  month = feb,
  journal = JSV,
  volume = {332},
  number = {4},
  pages = {714--724},
  issn = {0022-460X},
  doi = {10.1016/j.jsv.2012.10.014}
}

@article{levenberg-1944,
     doi = {10.1090/qam/10666},
     author = {K. Levenberg},
     journal = {Quarterly of Applied Mathematics},
     number = {2},
     pages = {164--168},
     publisher = {Brown University},
     title = {A method for the solution of certain non-linear problems in least squares},
     volume = {2},
     year = {1944}
}

@article{marquardt-1963,
    author = {Marquardt, Donald W.},
    title = {An Algorithm for Least-Squares Estimation of Nonlinear Parameters},
    journal = {Journal of the Society for Industrial and Applied Mathematics},
    volume = {11},
    number = {2},
    pages = {431-441},
    year = {1963},
    doi = {10.1137/0111030},
}

@article{guess1975Calculation,
  title = {Calculation of Perforated Plate Liner Parameters from Specified Acoustic Resistance and Reactance},
  author = {Guess, A. W.},
  year = {1975},
  month = may,
  journal = JSV,
  volume = {40},
  number = {1},
  pages = {119--137},
  publisher = {{Academic Press}},
  issn = {0022-460X},
  doi = {10.1016/S0022-460X(75)80234-3}
}

@article{ingard1959influence,
  title = {Influence of Fluid Motion Past a Plane Boundary on Sound Reflection, Absorption, and Transmission},
  author = {Ingard, Uno},
  year = {1959},
  journal = JASA,
  volume = {31},
  number = {7},
  pages = {1035--1036},
  issn = {0001-4966},
  doi = {10.1121/1.1907805}
}

@manual{ISO3966,
  type = {Standard},
  author = {{International Organization for Standardization}},
  title = {{ISO} 3966:2008. {Measurement} of Fluid Flow in Closed Conduits \textemdash{} {Velocity} Area Method Using {Pitot} Static Tubes},
  year = {2008},
  address = {{Genebra, Switzerland}},
  institution = {{International Organization for Standardization}},
  org-short = {ISO}
}

@article{jing2008straightforward,
  title = {A Straightforward Method for Wall Impedance Eduction in a Flow Duct},
  author = {Jing, Xiaodong and Peng, Sen and Sun, Xiaofeng},
  year = {2008},
  journal = JASA,
  volume = {124},
  number = {1},
  pages = {227--234},
  issn = {0001-4966},
  doi = {10.1121/1.2932256}
}

@article{jingInvestigationStraightforwardImpedance2015,
  title = {Investigation of Straightforward Impedance Eduction in the Presence of Shear Flow},
  author = {Jing, Xiaodong and Peng, Sen and Wang, Lixun and Sun, Xiaofeng},
  year = {2015},
  month = jan,
  journal = JSV,
  volume = {335},
  pages = {89--104},
  issn = {0022-460X},
  doi = {10.1016/j.jsv.2014.08.031}}

@article{yangShearFlowEffects2024,
  title = {Shear Flow Effects in a {{2D}} Duct: {{Influence}} on Wave Propagation and Direct Impedance Eduction},
  shorttitle = {Shear Flow Effects in a {{2D}} Duct},
  author = {Yang, Jinyue and Humbert, Thomas and Golliard, Joachim and Gabard, Gw{\'e}na{\"e}l},
  year = {2024},
  month = apr,
  journal = JSV,
  volume = {576},
  pages = {118296},
  issn = {0022-460X},
  doi = {10.1016/j.jsv.2024.118296},
  urldate = {2024-05-15}}

@article{don1997Accuracy,
  title = {Accuracy {{Enhancement}} for {{Higher Derivatives}} Using {{Chebyshev Collocation}} and a {{Mapping Technique}}},
  author = {Don, Wai Sun and Solomonoff, Alex},
  year = {1997},
  month = jul,
  journal = {SIAM Journal on Scientific Computing},
  volume = {18},
  number = {4},
  pages = {1040--1055},
  publisher = {{Society for Industrial and Applied Mathematics}},
  issn = {1064-8275},
  doi = {10.1137/S1064827594274607}
}

@article{khamis2017acoustics,
  title = {Acoustics in a Two-Deck Viscothermal Boundary Layer over an Impedance Surface},
  author = {Khamis, Doran and Brambley, Edward James},
  year = {2017},
  month = oct,
  journal = AIAAJ,
  volume = {55},
  number = {10},
  pages = {3328--3345},
  publisher = {{American Institute of Aeronautics and Astronautics}},
  issn = {0001-1452},
  doi = {10.2514/1.J055598}
}

@inproceedings{kooi1981Experimental,
  title = {An Experimental Study of the Acoustic Impedance of {{Helmholtz}} Resonator Arrays under a Turbulent Boundary Layer},
  booktitle = {7th {{Aeroacoustics Conference}}},
  author = {Kooi, J. and Sarin, S.},
  year = {1981},
  month = oct,
  publisher = {{American Institute of Aeronautics and Astronautics (AIAA)}},
  address = {{Palo Alto,CA,U.S.A.}},
  doi = {10.2514/6.1981-1998},
  pages={1998},
}

@inproceedings{murray2012development,
  title = {Development of a Single Degree of Freedom Perforate Impedance Model under Grazing Flow and High {{SPL}}},
  booktitle = {18th {{AIAA}}/{{CEAS}} Aeroacoustics Conference (33rd {{AIAA}} Aeroacoustics Conference)},
  author = {Murray, Paul and Astley, R. Jeremy},
  year = {2012},
  month = jun,
  publisher = {{American Institute of Aeronautics and Astronautics}},
  address = {{Colorado Springs, Colorado}},
  doi = {10.2514/6.2012-2294},
  isbn = {978-1-60086-932-7},
  pages={2294},
}

@article{myers1980acoustic,
  title = {On the Acoustic Boundary Condition in the Presence of Flow},
  author = {Myers, M K},
  year = {1980},
  journal = JSV,
  volume = {71},
  pages = {429--434},
  issn = {0022460X},
  doi = {10.1016/0022-460X(80)90424-1}
}

@article{nayfeh1974Effect,
  title = {Effect of Mean-Velocity Profile Shapes on Sound Transmission through Two-Dimensional Ducts},
  author = {Nayfeh, A. H. and Kaiser, J. E. and Shaker, B. S.},
  year = {1974},
  month = jun,
  journal = JSV,
  volume = {34},
  number = {3},
  pages = {413--423},
  issn = {0022-460X},
  doi = {10.1016/S0022-460X(74)80320-2}
}

@article{pridmorebrown1958sound,
  title = {Sound Propagation in a Fluid Flowing through an Attenuating Duct},
  author = {{Pridmore-Brown}, D. C.},
  year = {1958},
  month = aug,
  journal = JFM,
  volume = {4},
  number = {04},
  pages = {393--406},
  publisher = {{Cambridge University Press}},
  issn = {0022-1120},
  doi = {10.1017/S0022112058000537}
}

@inproceedings{primus2013ONERANASA,
  title = {{{ONERA-NASA}} Cooperative Effort on Liner Impedance Eduction},
  booktitle = {19th {AIAA/CEAS} Aeroacoustics Conference},
  author = {Primus, Julien and Piot, Estelle and Simon, Frank and Jones, Michael G. and Watson, Willie},
  year = {2013},
  pages = {2273},
  doi = {10.2514/6.2013-2273}
}

@article{renou2011Failure,
  title = {Failure of the {{Ingard}}\textendash{{Myers}} Boundary Condition for a Lined Duct: {{An}} Experimental Investigation},
  author = {Renou, Yga{\"a}l and Aur{\'e}gan, Yves},
  year = {2011},
  month = jul,
  journal = JASA,
  volume = {130},
  number = {1},
  pages = {52},
  publisher = {{Acoustical Society of AmericaASA}},
  issn = {0001-4966},
  doi = {10.1121/1.3586789}
}

@article{roncen2020WavenumberBased,
  title = {Wavenumber-{{Based Impedance Eduction}} with a {{Shear Grazing Flow}}},
  author = {Roncen, R. and Piot, E. and M{\'e}ry, F. and Simon, F. and Jones, M. G. and Nark, D. M.},
  year = {2020},
  journal = AIAAJ,
  volume = {58},
  number = {7},
  pages = {3040--3050},
  publisher = {{American Institute of Aeronautics and Astronautics}},
  issn = {0001-1452},
  doi = {10.2514/1.J059100}
}

@book{smith1989aircraft,
  title = {Aircraft Noise},
  author = {Smith, Michael J. T.},
  year = {1989},
  pages = {382},
  publisher = {{Cambridge University Press}},
  isbn = {978-0-511-58452-7}
}

@article{watson1999validation,
  title = {Validation of an Impedance Eduction Method in Flow},
  author = {Watson, W and Jones, M and Parrott, T},
  year = {1999},
  journal = AIAAJ,
  volume = {37},
  number = {7},
  pages = {818--824},
  doi = {10.2514/2.7529}
}

@article{weng2018Flow,
  title = {Flow and {{Viscous Effects}} on {{Impedance Eduction}}},
  author = {Weng, Chenyang and Schulz, Anita and Ronneberger, Dirk and Enghardt, Lars and Bake, Friedrich},
  year = {2018},
  month = mar,
  journal = AIAAJ,
  volume = {56},
  number = {3},
  pages = {1118--1132},
  publisher = {{American Institute of Aeronautics and Astronautics}},
  issn = {0001-1452},
  doi = {10.2514/1.J055838}
}

@article{boyer2011Theoretical,
  title = {Theoretical Investigation of Hydrodynamic Surface Mode in a Lined Duct with Sheared Flow and Comparison with Experiment},
  author = {Boyer, Germain and Piot, Estelle and Brazier, Jean-Philippe},
  year = {2011},
  month = apr,
  journal = JSV,
  volume = {330},
  number = {8},
  pages = {1793--1809},
  issn = {0022-460X},
  doi = {10.1016/j.jsv.2010.10.035}
}

@article{bonomoComparisonSituImpedance2024,
  title = {A Comparison of in Situ and Impedance Eduction Experimental Techniques for Acoustic Liners with Grazing Flow and High Sound Pressure Level},
  author = {Bonomo, Lucas A and Quintino, Nicolas T and Spillere, Andr{\'e} M N and Murray, Paul B and Cordioli, Julio A},
  year = {2024},
  month = jan,
  journal = IJA,
  pages = {Article in Advance},
  publisher = {{SAGE Publications}},
  issn = {1475-472X},
  doi = {10.1177/1475472X231225629}
}

@article{rashidi_3d_2025,
    title = {{3D} {Multimodal} inverse method for liner impedance eduction},
    issn = {0022-460X},
    url = {https://www.sciencedirect.com/science/article/pii/S0022460X25000288},
    doi = {10.1016/j.jsv.2025.118954},
    urldate = {2025-02-10},
    journal = {Journal of Sound and Vibration},
    author = {Rashidi, Hamid and Golliard, Joachim and Humbert, Thomas},
    month = jan,
    year = {2025},
    pages = {118954},
}

@incollection{brooks2007Sound,
  title = {Sound Transmission in Ducts with Sheared Mean Flow},
  booktitle = {13th {AIAA}/{CEAS} Aeroacoustics Conference},
  date = {2007-05-21},
  year={2007},
  pages={3545},
  author = {Brooks, Christopher and McAlpine, Alan},
  publisher = {{American Institute of Aeronautics and Astronautics}},
  doi = {10.2514/6.2007-3545}
}

@incollection{weng2018Comparison,
  title = {Comparison of {{Non-Modal-Based}} and {{Modal-Based Impedance Eduction Techniques}}},
  booktitle = {2018 {AIAA}/{CEAS} Aeroacoustics Conference},
  author = {Weng, Chenyang and Enghardt, Lars and Bake, Friedrich},
  date = {2018-06-24},
  year = {2018},
  pages = {3773},
  series = {{AIAA AVIATION Forum}},
  publisher = {{American Institute of Aeronautics and Astronautics}},
  doi = {10.2514/6.2018-3773}
}

\end{document}